\documentclass[reprint,noshowpacs,noshowkeys,prd,balancelastpage,nofootinbib]{revtex4}
\usepackage{eurosym}
\usepackage{subfig}
\usepackage{amsfonts}
\usepackage{mdframed}
\usepackage{amssymb}
\usepackage{footnote}
\usepackage{amsmath}
\usepackage{graphicx}
\usepackage{float}
\usepackage[font={footnotesize,it}]{caption}
\usepackage[utf8]{inputenc}
\usepackage{natbib}
\usepackage{tikz}
\usepackage{tikz-3dplot}
\usepackage[colorlinks=true,
            linkcolor=blue,
            urlcolor=blue,
            citecolor=blue]{hyperref}

\begin{document}

\title{Physical properties of charged black holes from the nonlinear
electrodynamics model based on electric potential regularization}
\author{S. Habib Mazharimousavi}
\email{habib.mazhari@emu.edu.tr}
\affiliation{Department of Physics, Faculty of Arts and Sciences, Eastern Mediterranean
University, Famagusta, North Cyprus via Mersin 10, T\"{u}rkiye}
\date{\today }

\begin{abstract}
We investigate static, spherically symmetric black hole solutions arising
from Einstein gravity minimally coupled to a nonlinear electrodynamics (NED)
model constructed from a regularized electric potential of a point charge.
The resulting spacetime is characterized by three parameters, namely, the ADM mass $%
M $, the electric charge $Q$, and the nonlinear scale $r_{0}$. We show that,
depending on the values of $M$ and $Q$, the solutions exhibit a rich causal
structure comprising black holes with single, double, and triple horizons,
as well as naked singular geometries. The nature of the central singularity
is determined by the combination $m-\frac{2}{3}q^{2}$, allowing for both
spacelike and timelike singularities. We perform a detailed thermodynamic
analysis by deriving the Hawking temperature and heat capacity, revealing
the existence of two critical charge parameters that govern the thermal
behavior. Below a critical charge, the black holes are thermally unstable,
whereas above it a stable phase emerges within a finite range of horizon
radii bounded by Davies points. For sufficiently large charge, extremal
configurations with vanishing temperature arise, further constraining the
stability region. We also investigate observational signatures by analyzing
null geodesics and black hole shadows, showing that nonlinear
electrodynamics corrections lead to noticeable deviations from the
Reissner-Nordstr\"{o}m geometry, including the possible absence of a photon
sphere beyond a critical charge. Our results highlight that nonlinear
electrodynamics significantly enriches the causal structure, thermodynamic
phase space, and dynamical response of charged black holes, providing
potentially observable deviations from the Reissner-Nordstr\"{o}m paradigm.
\end{abstract}

\maketitle

\section{Introduction}

Black holes constitute one of the most profound predictions of general
relativity, providing a unique arena in which gravity, quantum theory, and
high-energy physics intersect. The classical solutions of Einstein's field
equations, such as the Schwarzschild and Reissner-Nordstr\"{o}m (RN)
geometries, have been extensively studied and are known to possess curvature
singularities where the classical theory breaks down \cite%
{Schwarzschild1916,Reissner1916,Nordstrom1918,HawkingEllis1973}. The
resolution or regularization of such singularities remains a central problem
in gravitational physics and has motivated a wide variety of approaches,
including quantum gravity, modified gravity theories, and nonlinear matter
sources. Among these approaches, nonlinear electrodynamics (NED) has emerged
as a particularly compelling framework. Originally introduced by Born and
Infeld to remove divergences in the self-energy of point charges \cite%
{BornInfeld1934}, NED theories modify Maxwell's linear dynamics at high
field strengths while recovering it in the weak-field limit. When coupled to
gravity, such theories can significantly alter the structure of black hole
solutions. A remarkable result in this direction is the regular Bardeen
black hole, later interpreted as a solution sourced by a nonlinear magnetic
monopole \cite{Bardeen1968,AyonBeatoGarcia2000}. Since then, a large class
of regular and singular black hole solutions supported by NED has been
constructed and analyzed, revealing rich phenomenology in both the geometric
and thermodynamic sectors \cite%
{AyonBeatoGarcia1998,Bronnikov2001,Dymnikova2004,BalartVagenas2014,
FanWang2016,Kruglov2017,Hendi2017}.

A key feature of NED-coupled gravity is its ability to regularize the
electromagnetic field at short distances, thereby modifying the near-origin
behavior of the spacetime. In many known models, this leads to regular black
holes with de Sitter cores or softened singularities \cite%
{Hayward2006,BambiModesto2013}. However, the precise nature of the central
structure - whether regular, spacelike singular, or timelike singular-
depends sensitively on the choice of the NED Lagrangian. This motivates the
systematic construction of NED models based on physically well-motivated
regular electric or magnetic configurations. In this context, a particularly
natural starting point is a regularized electric potential for a point
charge. Recently, in \cite{Mazharimousavi2026}, a NED model was constructed
from such a regular electric potential, leading to a well-defined and finite
electric field everywhere in spacetime. Unlike many existing models that are
introduced at the level of the Lagrangian, this approach is grounded
directly in a physically regular field configuration. The resulting NED
Lagrangian exhibits the correct Maxwellian limit at large distances while
encoding nonlinear corrections that become significant near the origin. In 
\cite{Mazharimousavi2026}, we coupled this NED model to Einstein gravity and
derived the resulting static, spherically symmetric solutions. The metric
function obtained in this framework generalizes the standard RN geometry by
incorporating a nonlinear scale parameter $r_{0}$, which controls the
strength of the NED corrections. The resulting spacetime is characterized by
three independent parameters that are the ADM mass $M$, the electric charge $%
Q$, and the nonlinear scale $r_{0}$. The latter additional degree of freedom
leads to a significantly richer structure compared to the linear
electrovacuum case.

Following our initial work on this NED model, here in this current study, we
study further the physical properties of the solutions. One of the central
aspects of our analysis is the classification of the global structure of the
solutions. Depending on the values of the dimensionless parameters $%
m=M/r_{0} $ and $q=Q/r_{0}$, the spacetime admits multiple qualitatively
distinct configurations, including black holes with non-degenerate,
double-degenerate (extremal), and even triple-degenerate (ultra-extremal)
horizons, as well as naked singular geometries. The nature of the central
singularity is shown to depend on the combination $m-\frac{2}{3}q^{2}$,
leading to either spacelike or timelike behavior. Such a rich causal
structure goes beyond the standard RN case and highlights the nontrivial
role of nonlinear electromagnetic effects.

In addition to the geometric properties, we perform a detailed thermal
stability analysis of the solutions. Black hole thermodynamics has provided
deep insights into the interplay between gravity, quantum theory, and
statistical physics \cite{Bekenstein1973,Hawking1975}. In particular, the
stability of black holes can be studied through the behavior of the heat
capacity and the Hawking temperature. It is well known that the
Schwarzschild black hole is thermodynamically unstable, whereas charged
black holes can exhibit more intricate phase structures \cite%
{Davies1977,Chamblin1999}. In the present model, we show that the NED
corrections introduce two critical values of the charge parameter, leading
to the emergence of thermodynamically stable phases within finite regions of
parameter space.

Beyond thermodynamics, observational signatures of black holes have become
increasingly important with the advent of gravitational-wave astronomy and
black hole imaging. The shadow of a black hole, determined by unstable
circular null geodesics, provides a direct probe of the near-horizon
geometry \cite{Perlick2015,EHT2019a,EHT2019b}. Similarly, quasinormal modes
(QNMs), which describe the ringdown phase of perturbed black holes, encode
information about the underlying spacetime geometry \cite%
{ReggeWheeler1957,Zerilli1970,KokkotasSchmidt1999, BertiCardosoStarinets2009}%
. In the eikonal limit, QNMs are closely related to the properties of the
photon sphere \cite{Cardoso2009}, establishing a direct link between
geometric optics and wave dynamics.

Motivated by these developments, we extend our analysis to include the
shadow of the black hole solutions. We demonstrate that the nonlinear
electrodynamics corrections lead to systematic deviations from the RN case,
affecting the size of the shadow. These deviations become more pronounced in
the strong-field regime and may provide potential observational signatures
of nonlinear electromagnetic effects in gravitational systems.

The paper is organized as follows. In Sec.~II, we briefly review the NED
model and derive the corresponding black hole solutions. In Sec.~III, we
analyze the global causal structure and classify the possible
configurations. Sec.~IV is devoted to the thermodynamic properties and
stability analysis. In Sec.~V, we study the black hole shadow and its
dependence on the model parameters. Finally, we summarize our results and
discuss possible future directions in Sec.~VI.

\section{A quick review of the solutions}

In accordance with the NED model introduced in \cite{Mazharimousavi2026},
the regular electric potential of a point charge is given by 
\begin{equation}
\phi = \frac{Q}{\sqrt{r_{0}^{2}+r^{2}}},
\end{equation}
where $Q$ is the electric charge and $r_{0}$ is the NED parameter. The
corresponding electric field is also regular and is obtained as 
\begin{equation}
E = \frac{Qr}{\left( r_{0}^{2}+r^{2}\right) ^{3/2}}.
\end{equation}
The hybrid form of the NED Lagrangian is therefore obtained as 
\begin{equation}
\mathcal{L} = \frac{4Q^{2}}{r_{0}^{4}}\left[ 1-\frac{1+6\left( \frac{r^{2}}{%
r_{0}^{2}}\right) +4\left( \frac{r^{2}}{r_{0}^{2}}\right) ^{2}}{4\left( 
\frac{r}{r_{0}}\right) \left( 1+\frac{r^{2}}{r_{0}^{2}}\right) ^{3/2}}\right]%
,
\end{equation}
with 
\begin{equation}
\mathcal{F} = -\frac{Q^{2}}{2r_{0}^{4}}\frac{\left( \frac{r^{2}}{r_{0}^{2}}%
\right)}{\left( 1+\frac{r^{2}}{r_{0}^{2}}\right) ^{3}}.
\end{equation}

After coupling this NED model to gravity, the static spherically symmetric
metric is described by the line element 
\begin{equation}
ds^{2}=-f(r)\,dt^{2}+\frac{1}{f(r)}\,dr^{2}+r^{2}\left( d\theta ^{2}+\sin
^{2}\theta \,d\phi ^{2}\right) ,  \label{g}
\end{equation}%
where the metric function is given by 
\begin{equation}
f(r)=1-\frac{2M}{r}+\frac{8Q^{2}}{3r_{0}^{2}}\left( \frac{r^{2}}{r_{0}^{2}}-%
\frac{\frac{r^{2}}{r_{0}^{2}}-\frac{1}{2}}{\frac{r}{r_{0}}}\sqrt{1+\frac{%
r^{2}}{r_{0}^{2}}}\right) ,  \label{MF}
\end{equation}%
in which $M$ is the ADM mass of the black hole solution. Introducing the
dimensionless radial coordinate $x=\frac{r}{r_{0}}$, the dimensionless ADM
mass $m=\frac{M}{r_{0}}$, and the dimensionless electric charge $q=\frac{Q}{%
r_{0}}$, the metric function reduces to 
\begin{equation}
f(x)=1-\frac{2m}{x}+\frac{8q^{2}}{3}\left( x^{2}-\frac{2x^{2}-1}{2x}\sqrt{%
1+x^{2}}\right) .  \label{MF2}
\end{equation}%
From (\ref{MF2}), the asymptotic behavior of the metric function is given by 
\begin{equation}
\lim_{x\rightarrow \infty }f(x)\rightarrow 1-\frac{2m}{x}+\frac{q^{2}}{x^{2}}%
+\mathcal{O}\left( x^{-4}\right) ,
\end{equation}%
while the behavior for small $x$ reads 
\begin{equation}
\lim_{x\rightarrow 0}f(x)\rightarrow 1-\frac{2m-\frac{4}{3}q^{2}}{x}+%
\mathcal{O}(x).
\end{equation}

The solution is singular at $x=0$, however, the nature of the singularity,
whether timelike or spacelike, strongly depends on the sign of $m-\frac{2}{3}%
q^{2}$. Specifically, if $m-\frac{2}{3}q^{2}\geq 0$, the singularity at the
origin is spacelike, whereas if $m-\frac{2}{3}q^{2}<0$, it is timelike.
Accordingly, the following four different configurations can be identified
based on the values of $q^{2}$ and $m$, \textbf{A}) $m<m_{\min }$, \textbf{B}%
) $m=m_{\min }$, \textbf{C}) $m_{\min }<m<m_{c}$, and \textbf{D}) $m_{c}<m$,
where 
\begin{equation}
m_{\min }=\frac{(2\sqrt{2}+1)\sqrt{\sqrt{2}-1}}{3},
\end{equation}%
and 
\begin{equation}
m_{c}=\frac{\varepsilon ^{5}-30\varepsilon ^{4}-132\varepsilon
^{3}-288\varepsilon -864}{18\left( \varepsilon ^{5}+4\varepsilon
^{4}-72\varepsilon ^{3}-576\right) }\approx 1.1947,\quad \text{with }%
\varepsilon =\sqrt[3]{12\left( 9+\sqrt{93}\right) }.
\end{equation}%
In the following sections, we study in detail the causal structure of each
configuration and their thermal stability.

\subsection{ $m<m_{\min }$,}

\begin{figure}[h!]
\centering
\subfloat[]{\includegraphics[width=3cm]{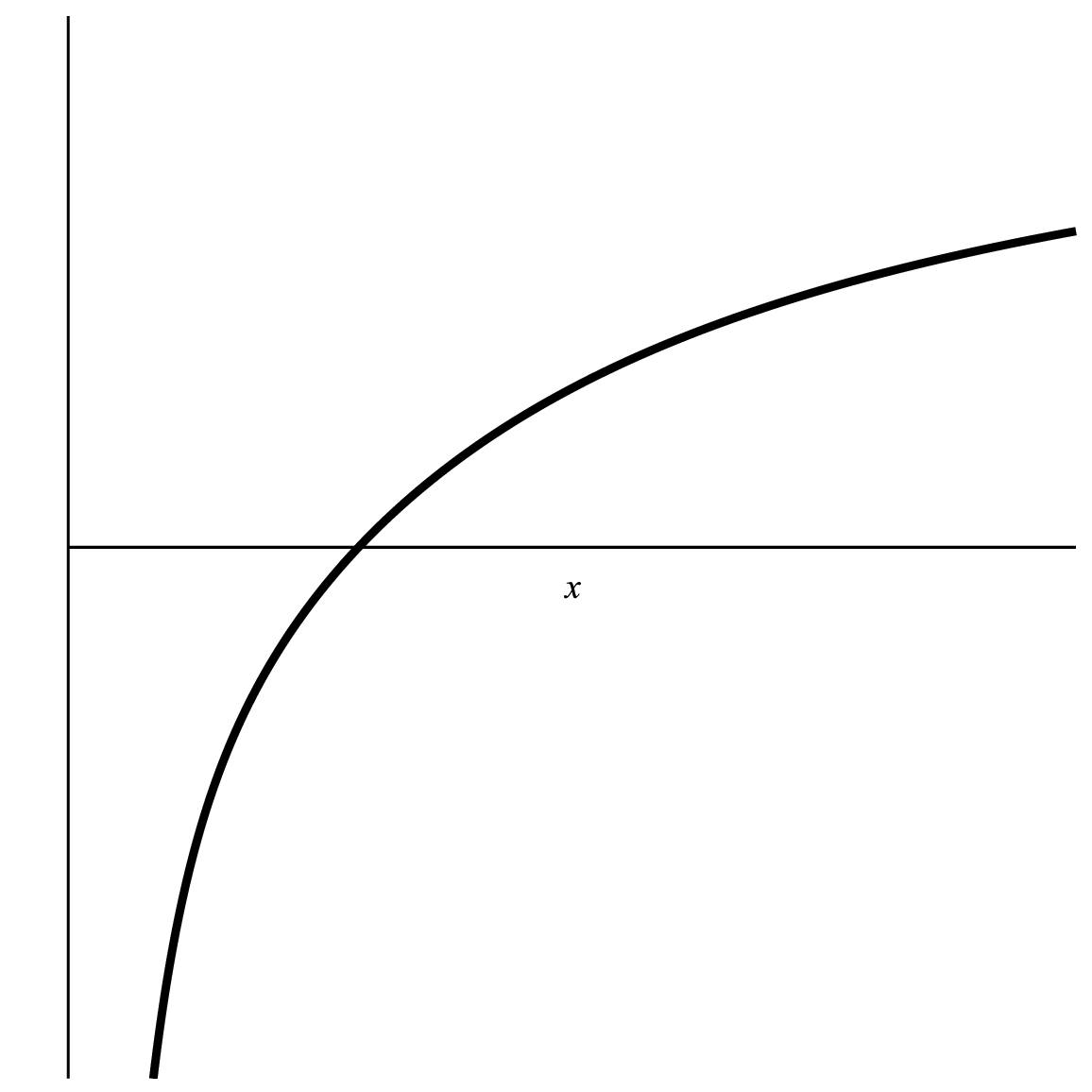}} \qquad \subfloat[]{%
\includegraphics[width=3cm]{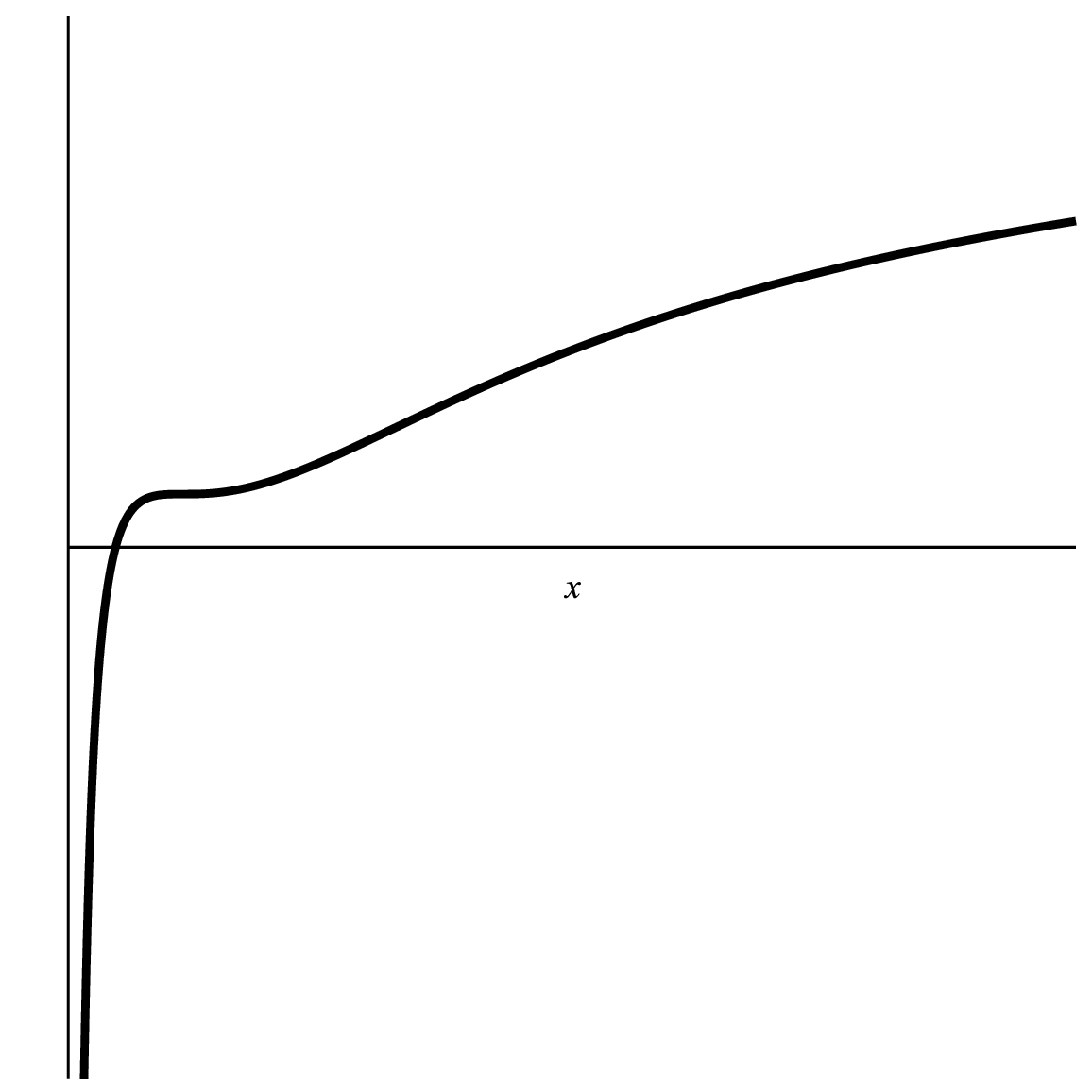}} \qquad \subfloat[]{%
\includegraphics[width=3cm]{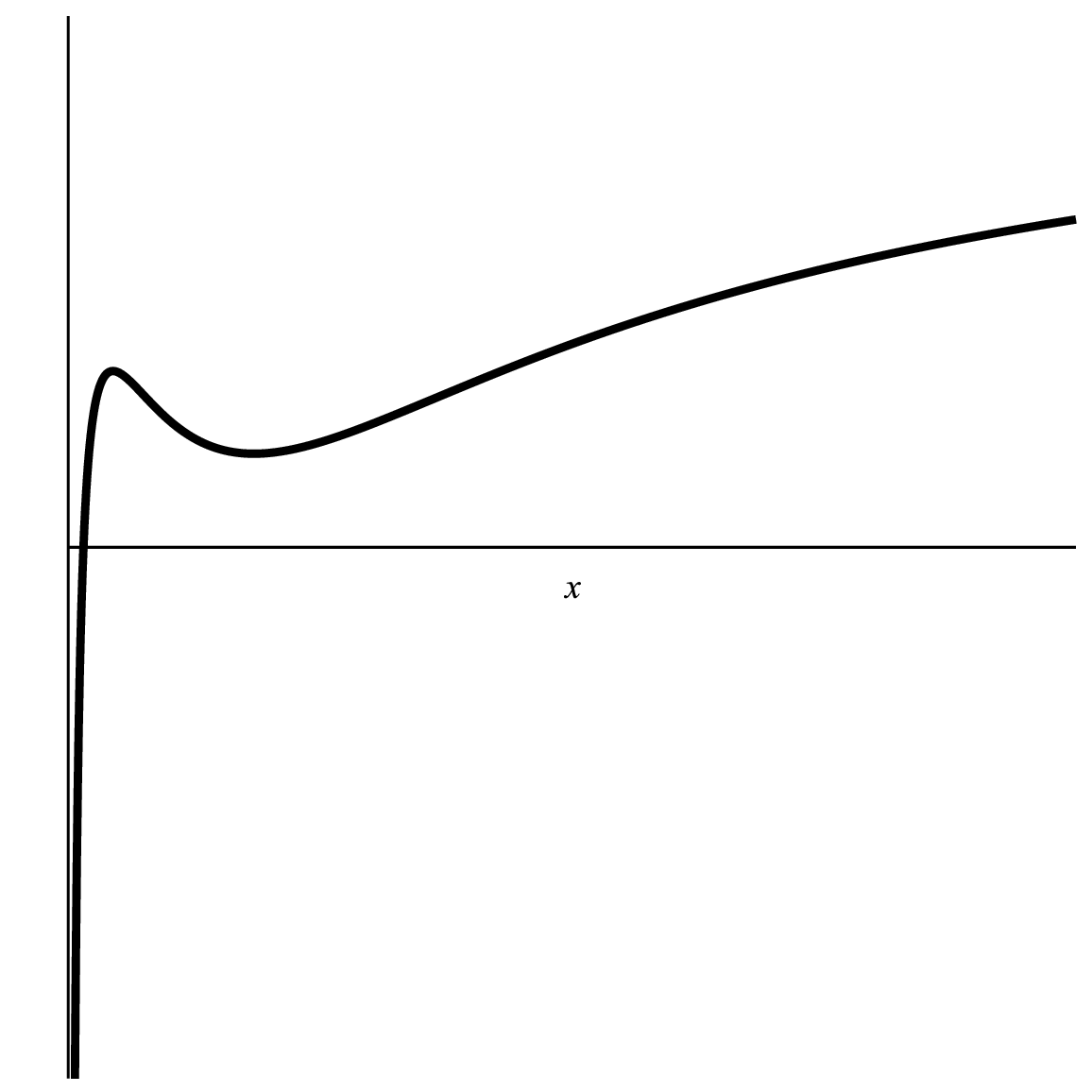}} \qquad \subfloat[]{%
\includegraphics[width=3cm]{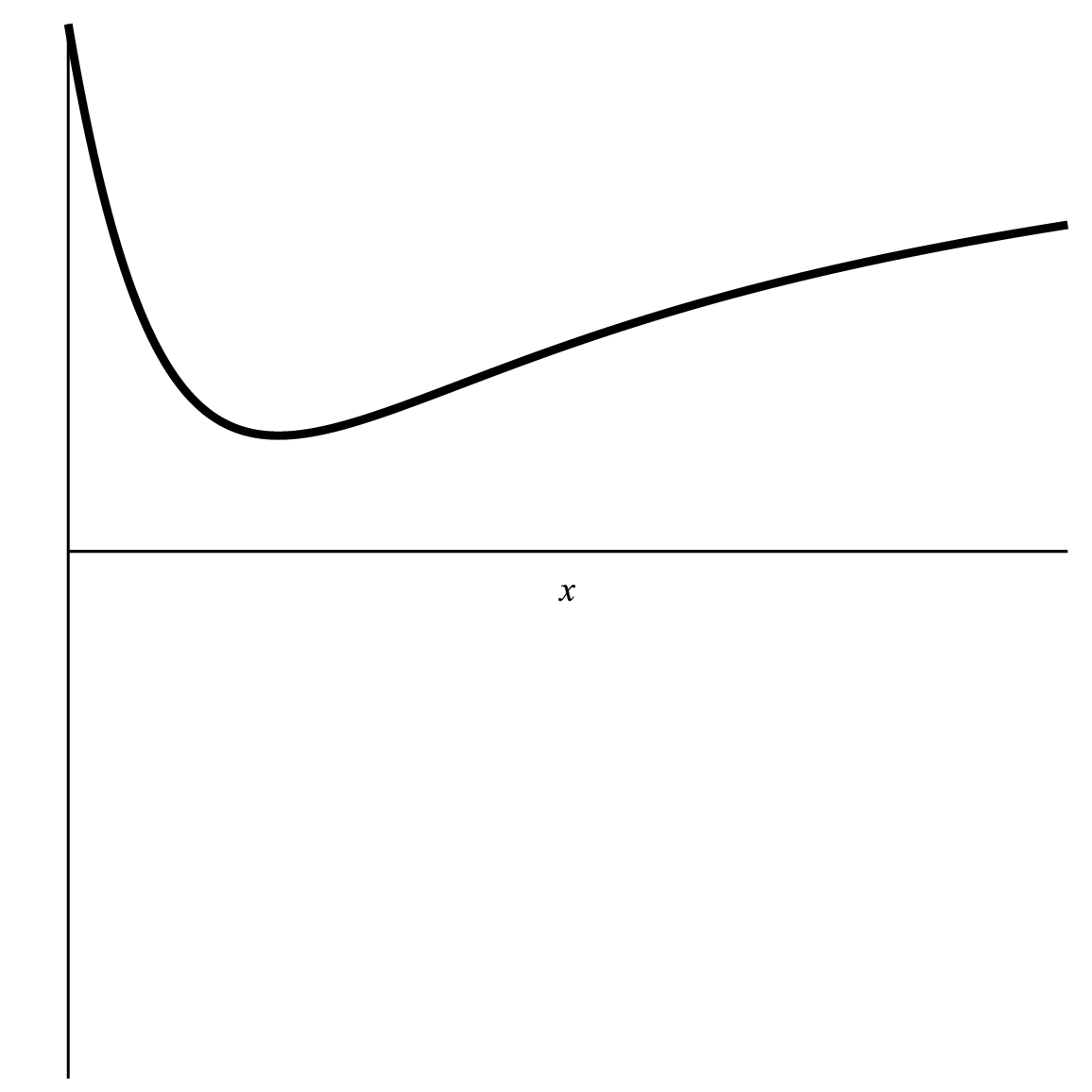}} \qquad \subfloat[]{%
\includegraphics[width=3cm]{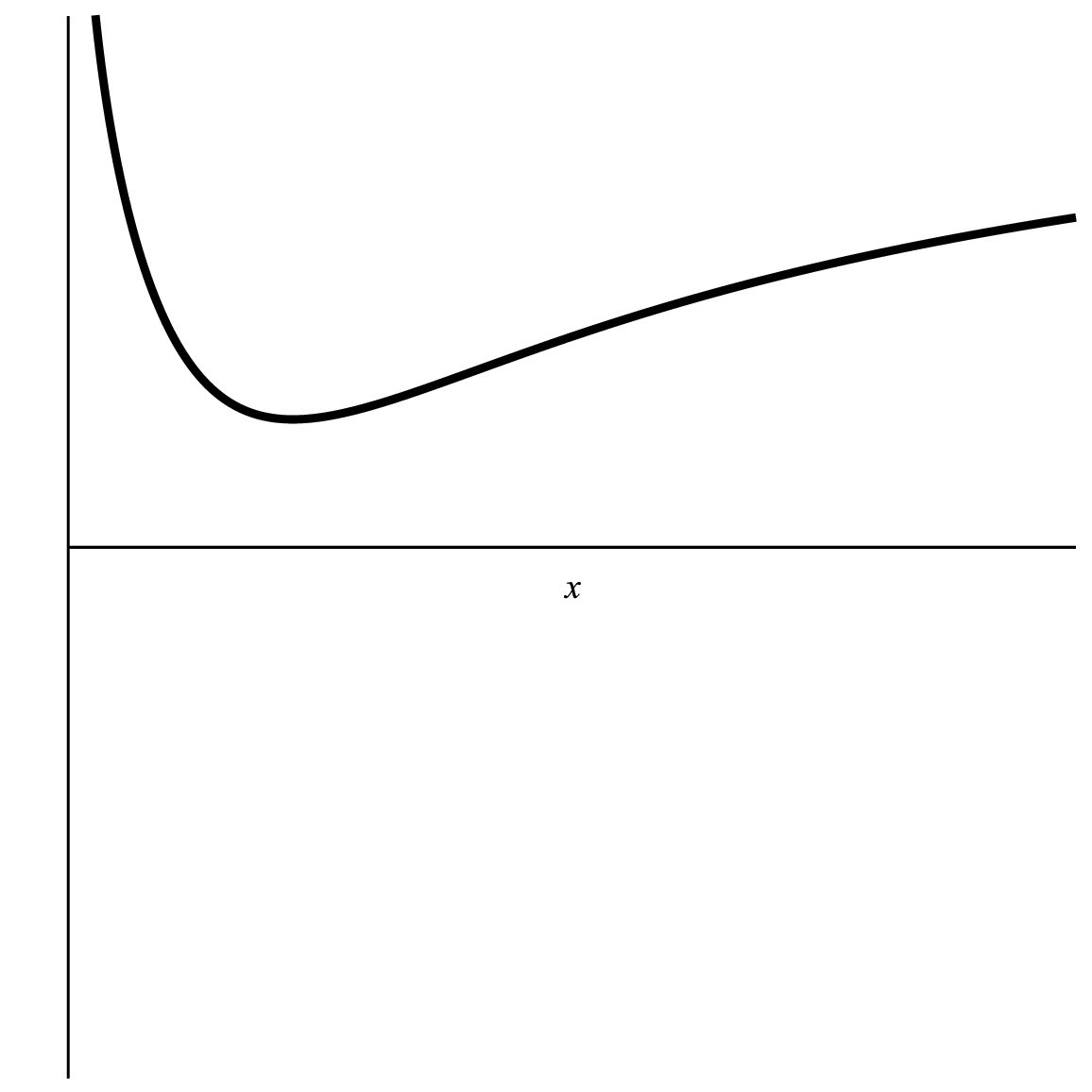}}
\caption{The metric function in Configuration 1, where $m<m_{\min}$ and $%
q^{2}$ increases from (a) to (e), with $q^{2}=\frac{3}{2}m$ in (d). Panels
(a), (b), and (c) admit an event horizon that hides the spacelike singularity at $r=0$, 
whereas panels (d) and (e) represent naked timelike singular solutions.}
\label{F1}
\end{figure}

\begin{figure}[h!]
\centering
\subfloat[]{\includegraphics[width=12cm]{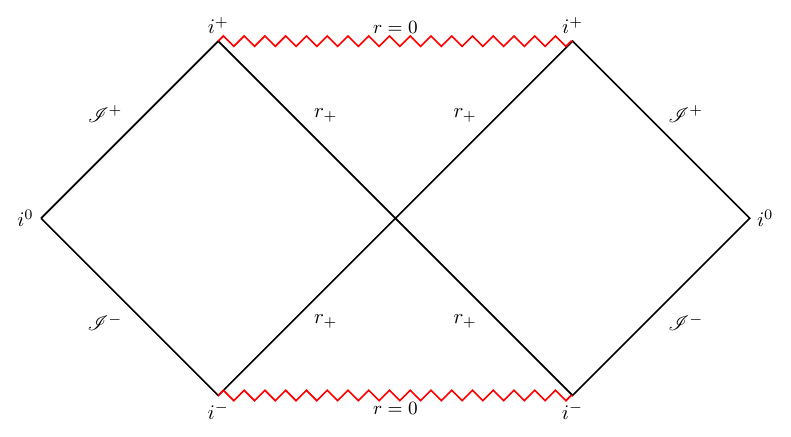}} \qquad %
\subfloat[]{\includegraphics[width=4cm]{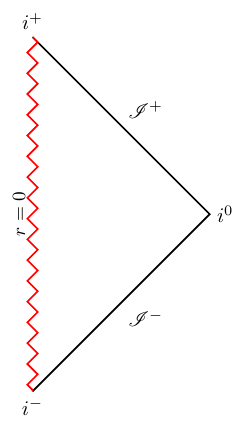}}
\caption{The Penrose diagrams of the solutions in Configurations 1 and 2.
Panel (a) corresponds to black hole solutions in Figs.~\protect\ref{F1} and 
\protect\ref{F2}, while panel (b) represents naked singular solutions in
Figs.~\protect\ref{F1} and \protect\ref{F2}. The singularity is spacelike
for black holes and timelike for non-black-hole solutions, as expected.
Moreover, panel (a) also represents the Penrose diagrams of Figs.~\protect
\ref{F3}(a) and \protect\ref{F4}(a).}
\label{P1}
\end{figure}

In Fig.~\ref{F1}, we plot the metric function for $m<m_{\min }$ and four
different values of $q^{2}$. From a horizon (i.e., $f(x_{h})=0$)
perspective, two distinct cases emerge. For $q^{2}<\frac{3}{2}m$, the
solution admits an event horizon such that the singularity becomes spacelike
and is hidden behind the horizon (see Fig.~\ref{F1}(a)-(c)). Figures~\ref{F1}%
(d) and (e) represent non-black-hole cases. Moreover, although the solutions
remain singular, the singularity is timelike. In particular, Fig.~\ref{F1}%
(d) corresponds to the case $m-\frac{2}{3}q^{2}=0$. In Fig.~\ref{P1}(a) and
(b), we present the Penrose diagrams for these two cases, where the nature
of the singularity is highlighted.

\subsection{ $m=m_{\min }$,}

\begin{figure}[h!]
\centering
\subfloat[]{\includegraphics[width=3cm]{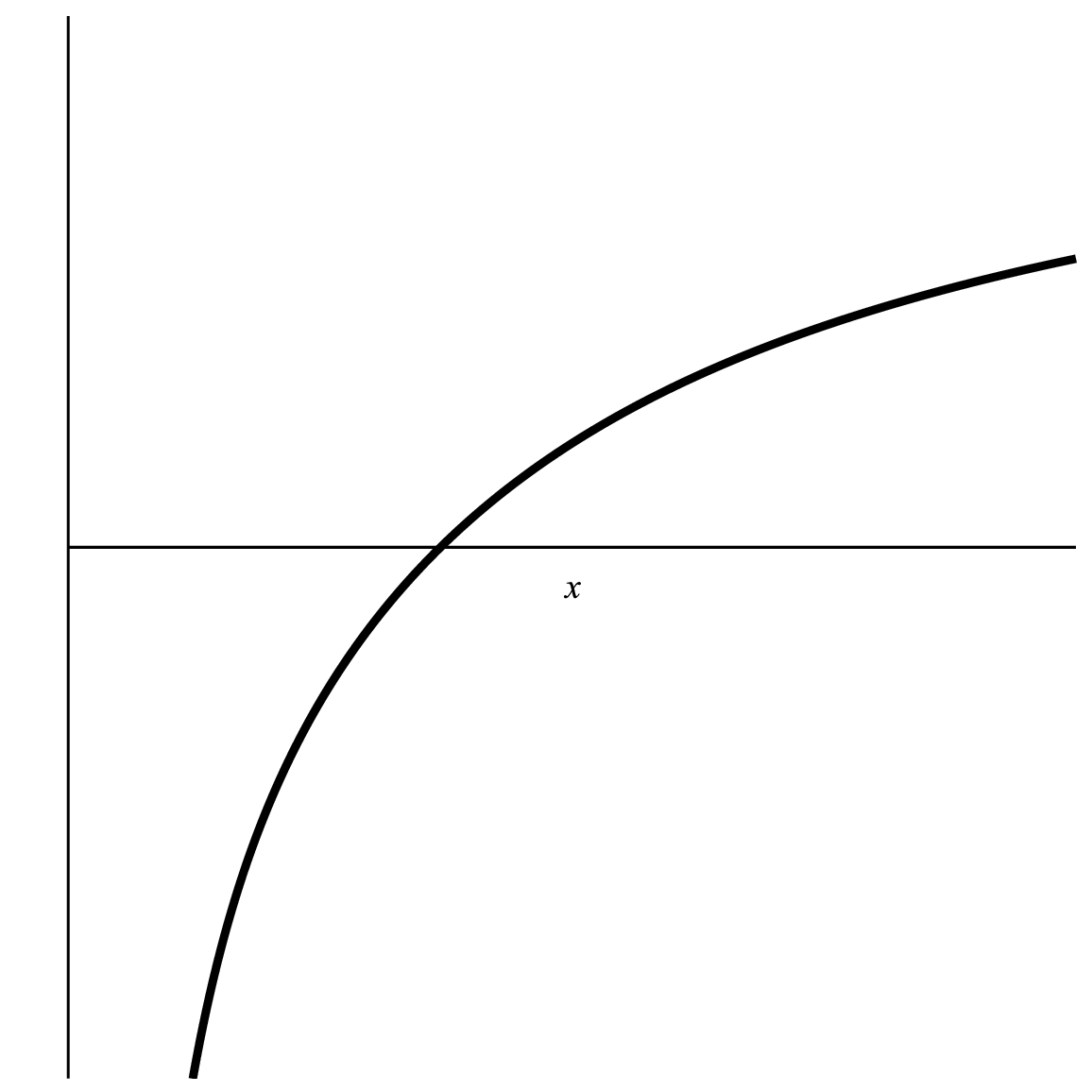}} \qquad \subfloat[]{%
\includegraphics[width=3cm]{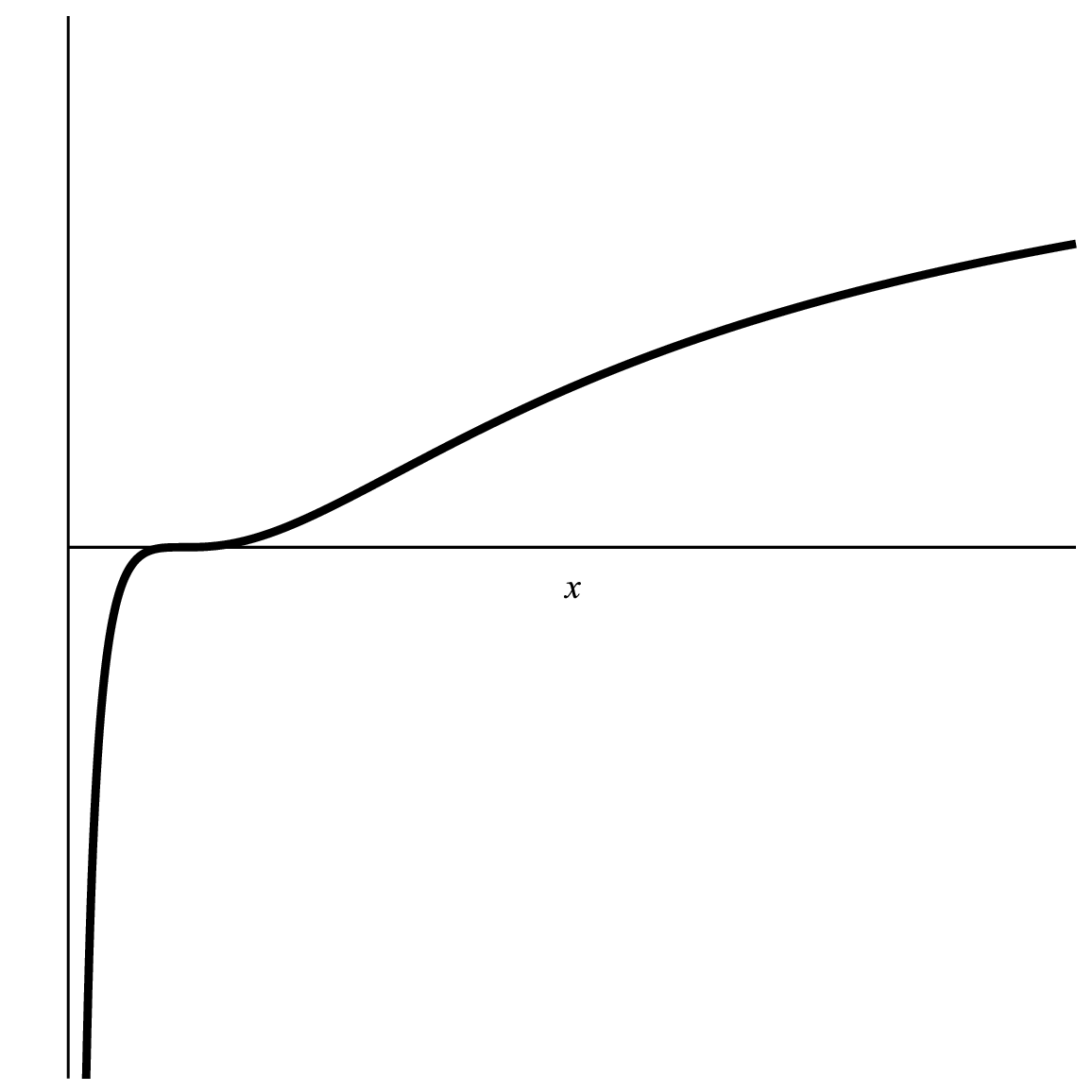}} \qquad \subfloat[]{%
\includegraphics[width=3cm]{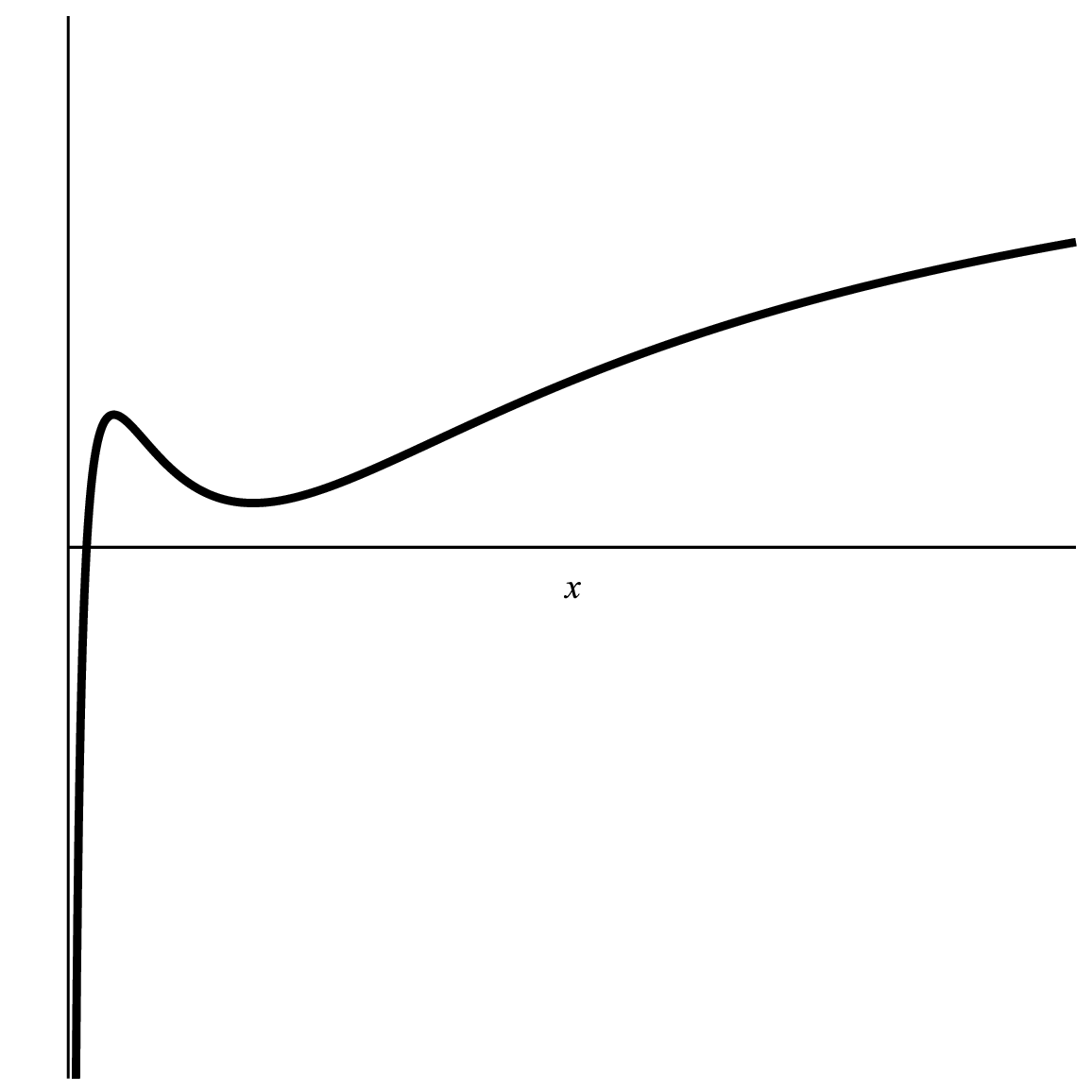}} \qquad \subfloat[]{%
\includegraphics[width=3cm]{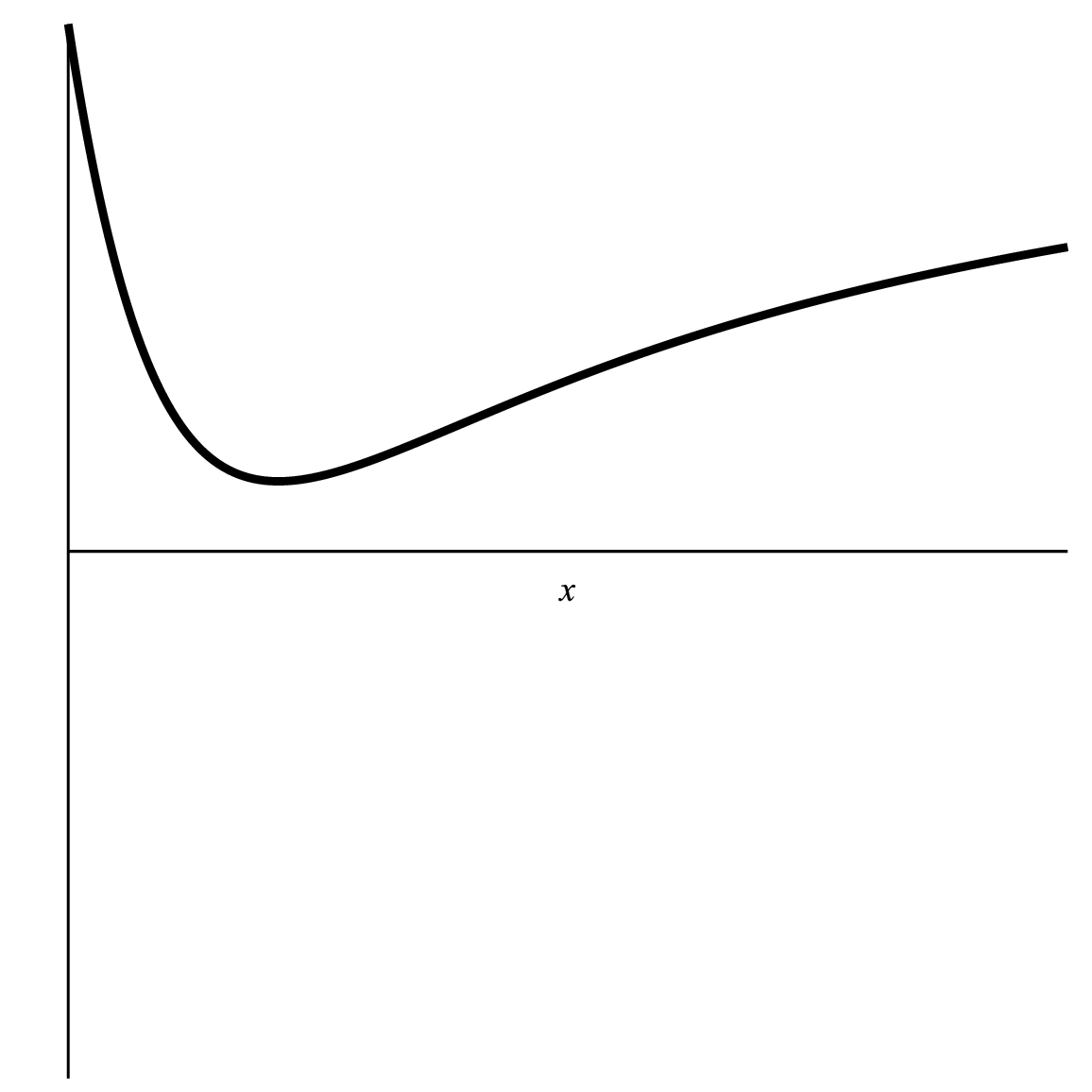}} \qquad \subfloat[]{%
\includegraphics[width=3cm]{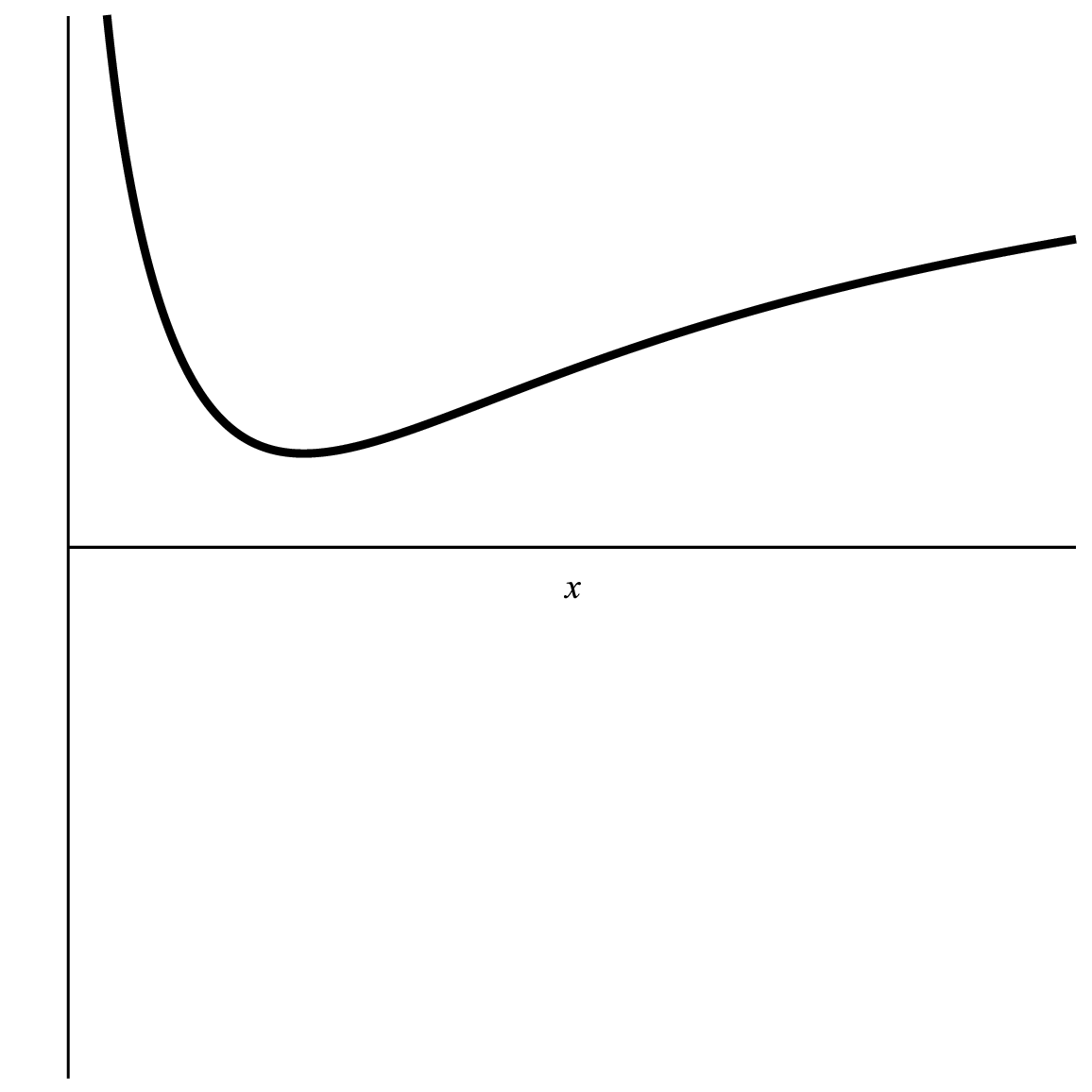}}
\caption{The metric function in Configuration 2, where $m=m_{\min}$ and $%
q^{2}$ increases from (a) to (e), with $q^{2}=\frac{3}{2}m$ in (d). Panels
(a), (b), and (c) admit an event horizon that hides the spacelike singularity at $r=0$%
, whereas panels (d) and (e) represent naked timelike singular solutions. In panel
(b), the metric function admits a triple horizon, i.e., $f=f^{\prime}=f^{%
\prime\prime}=0$ at $r=r_{+}$.}
\label{F2}
\end{figure}

The solutions in the second configuration are depicted in Fig.~\ref{F2} with 
$m=m_{\min}$ and increasing $q^{2}$ from (a) to (e). The value of $q^{2}$ in
(b) is $q_{\min}^{2}=\frac{3+2\sqrt{2}}{4}$, for which $f^{\prime}(x_{+})=0$
and therefore the event horizon is degenerate. Moreover, since $%
f^{\prime\prime}(x_{+})=0$ as well, the event horizon is a triple horizon.
Similar to Fig.~\ref{F1}, the metric functions in panels (d) and (e)
represent non-black-hole spacetimes with a naked singular center. Figure~\ref%
{F2}(d) corresponds to $q^{2}=\frac{3m}{2}$, at which the nature of the
singularity changes from spacelike for $q^{2}<\frac{3m}{2}$ to timelike for $%
q^{2}\geq \frac{3m}{2}$. This is illustrated in the Penrose diagrams shown
in Fig.~\ref{P1}, where panel (a) corresponds to the spacelike singularity
of black hole solutions (Fig.~\ref{F2}(a)-(c)), while panel (b) represents
the timelike singularity of naked singular solutions (Fig.~\ref{F2}(d) and
(e)).

\subsection{ $m_{\min }<m<m_{c}$,}

\begin{figure}[h!]
\centering
\subfloat[]{\includegraphics[width=2.4cm]{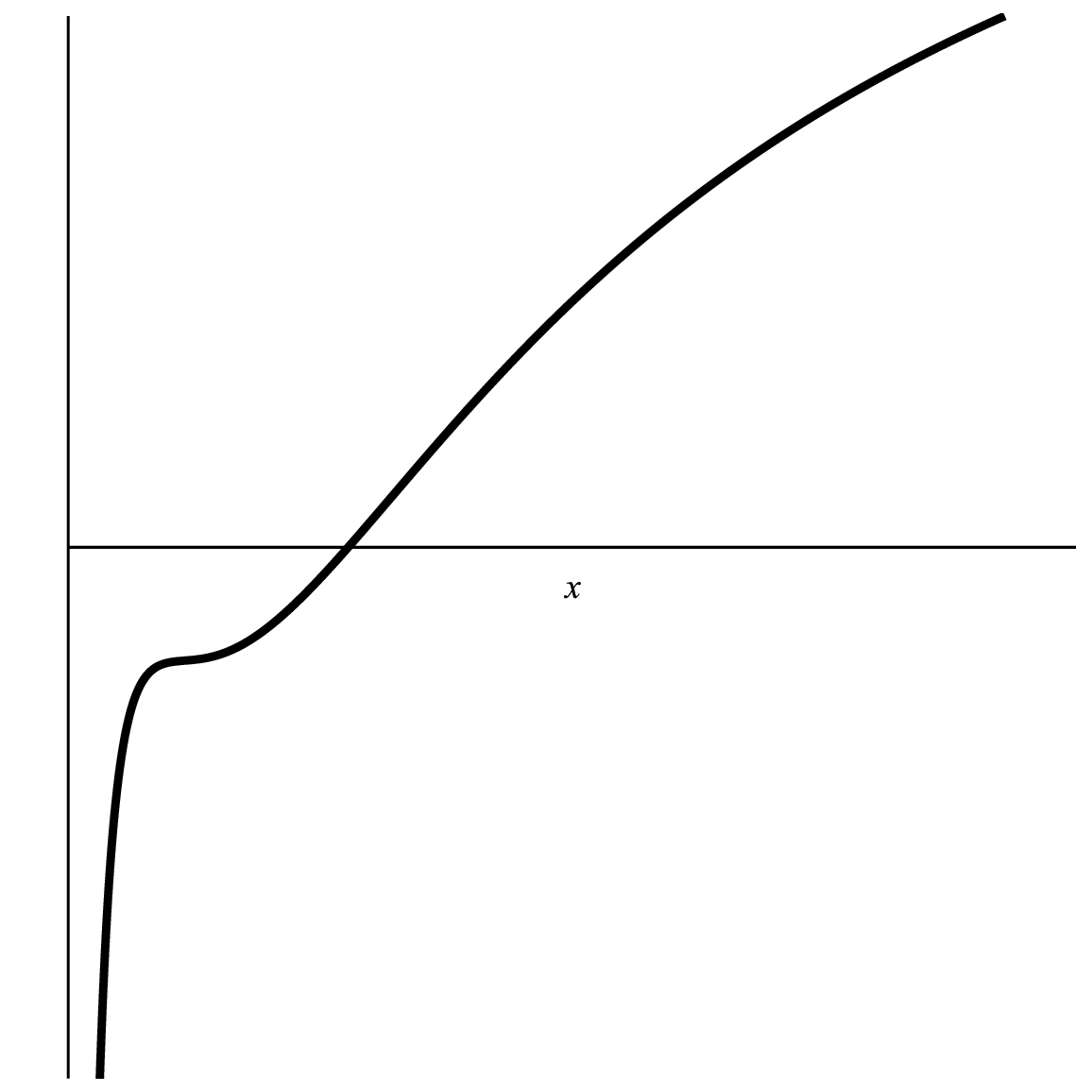}} \qquad %
\subfloat[]{\includegraphics[width=2.4cm]{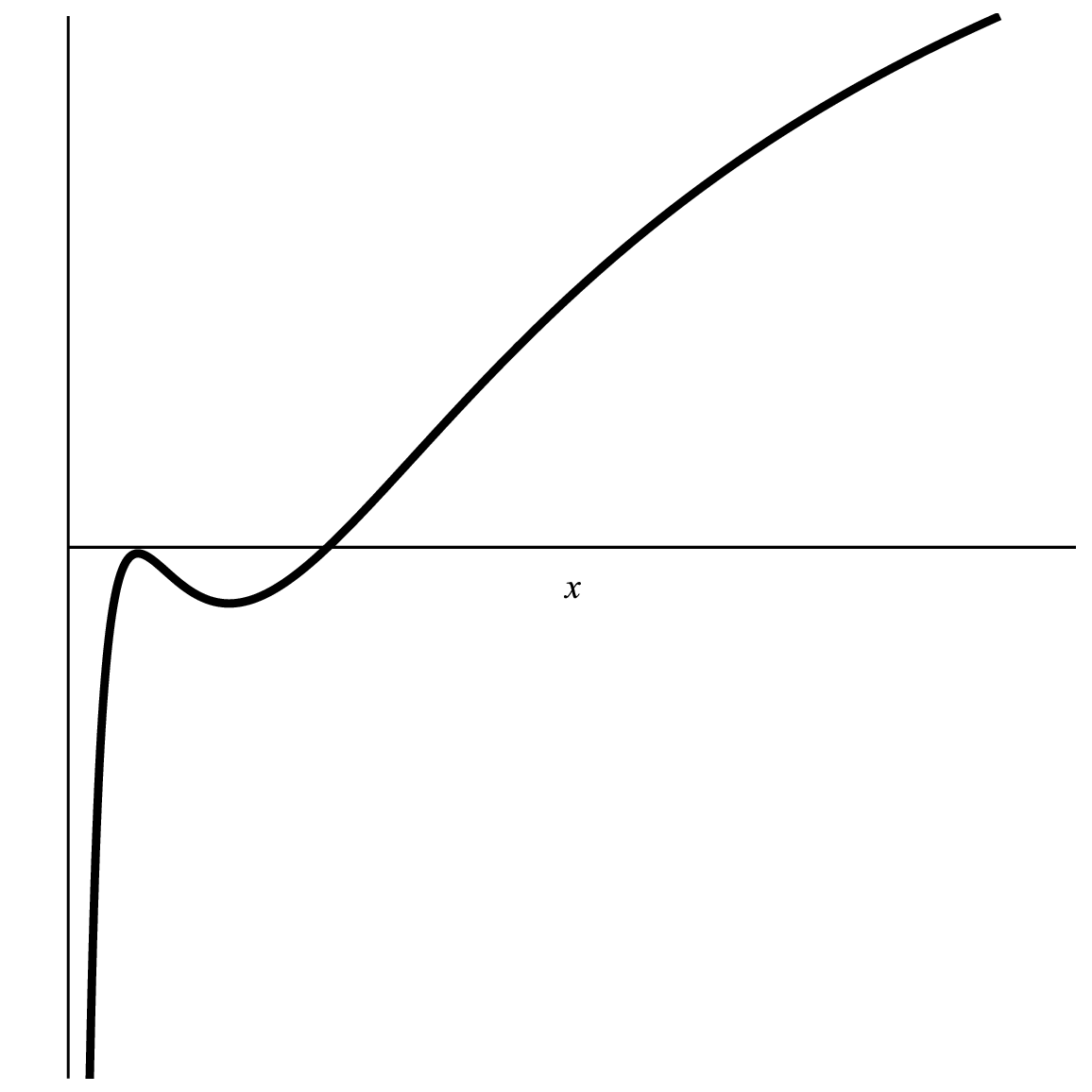}} \qquad %
\subfloat[]{\includegraphics[width=2.4cm]{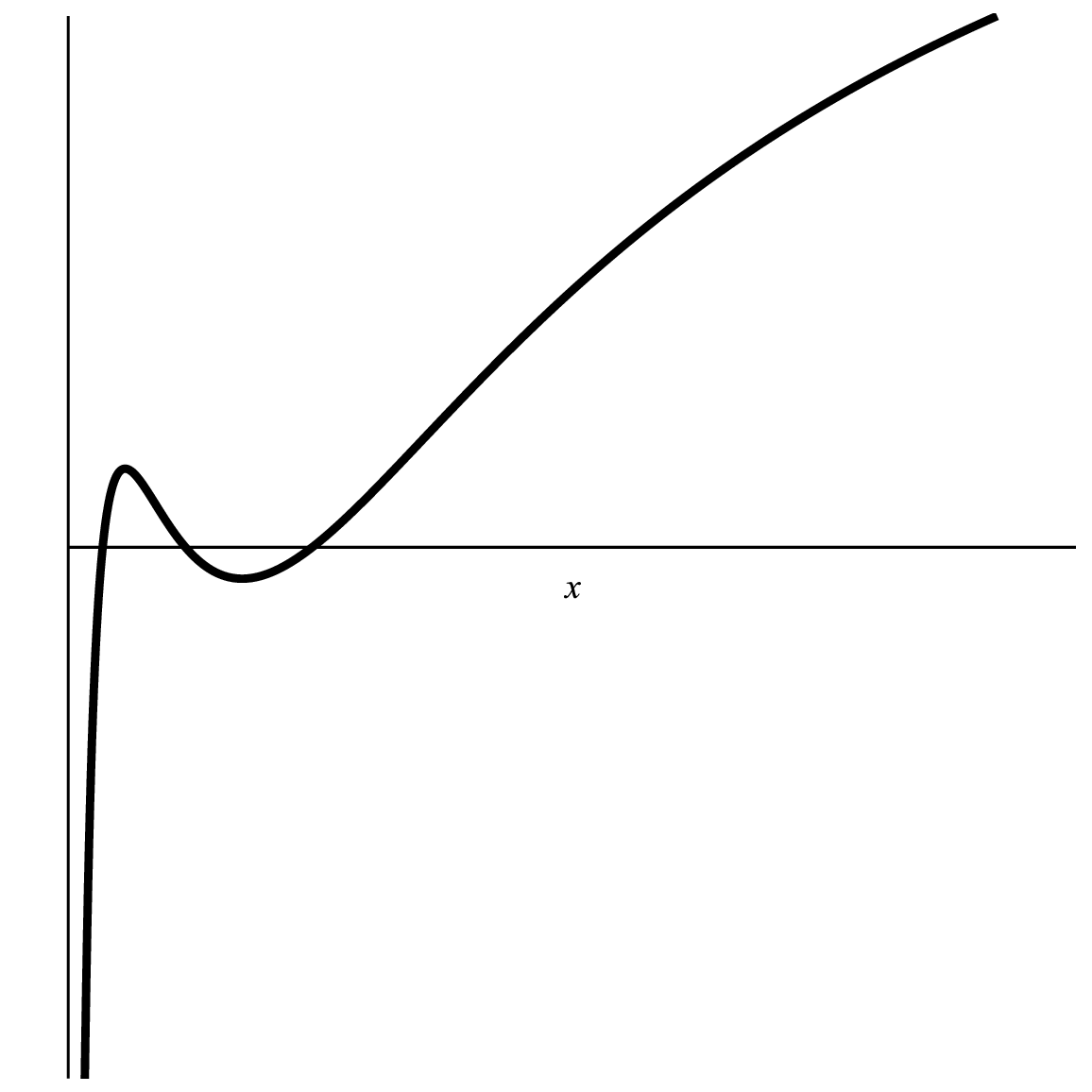}} \qquad %
\subfloat[]{\includegraphics[width=2.4cm]{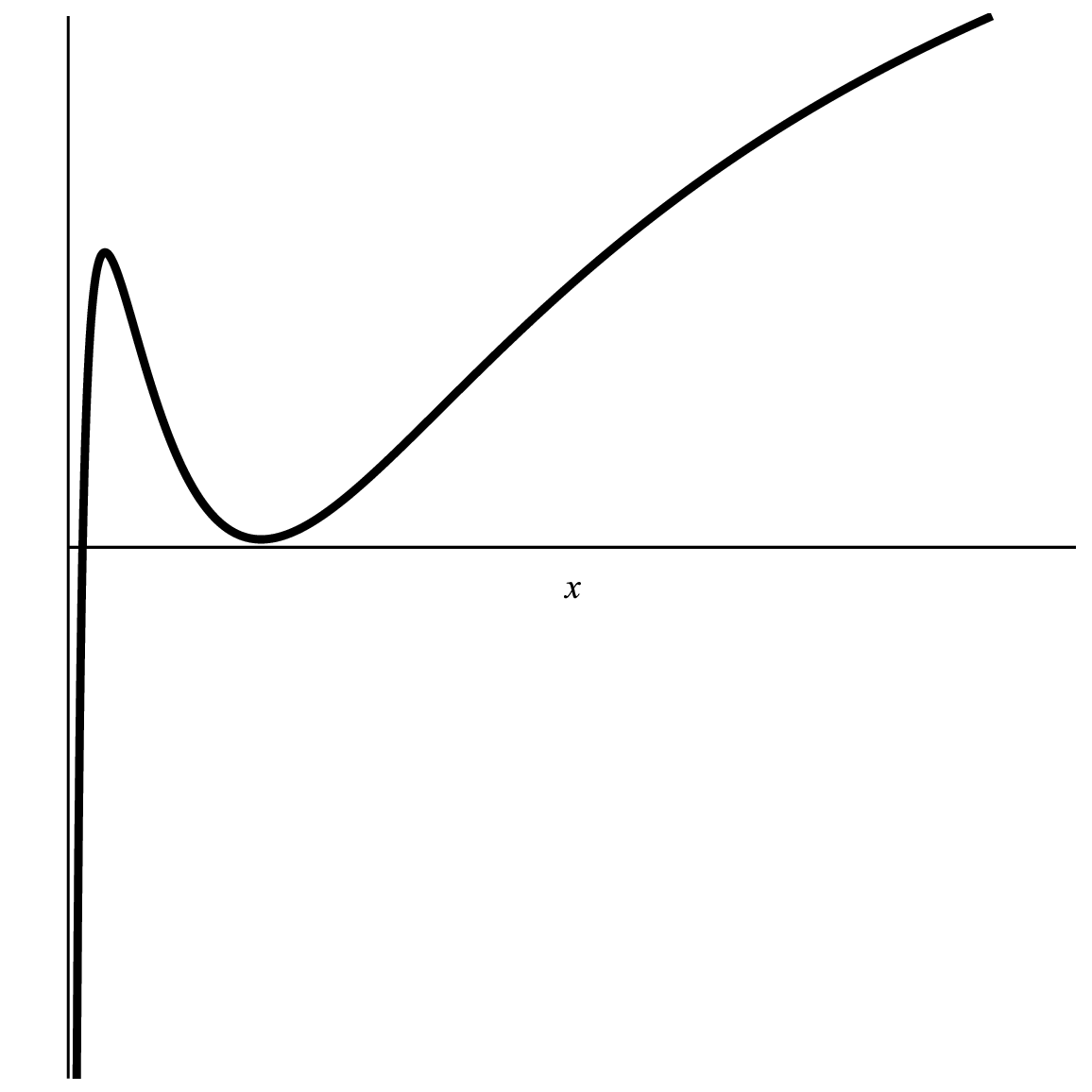}} \qquad %
\subfloat[]{\includegraphics[width=2.4cm]{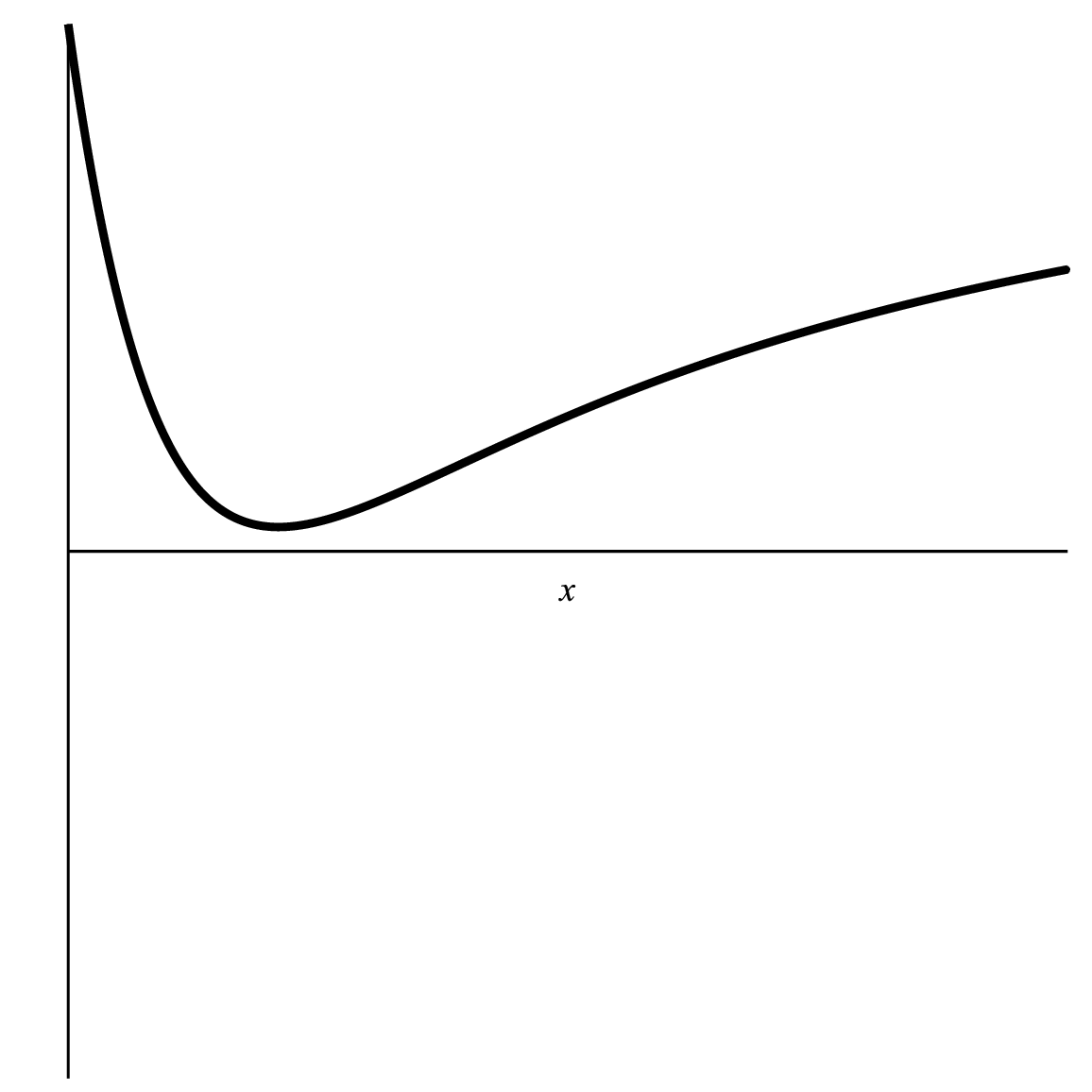}} \qquad %
\subfloat[]{\includegraphics[width=2.4cm]{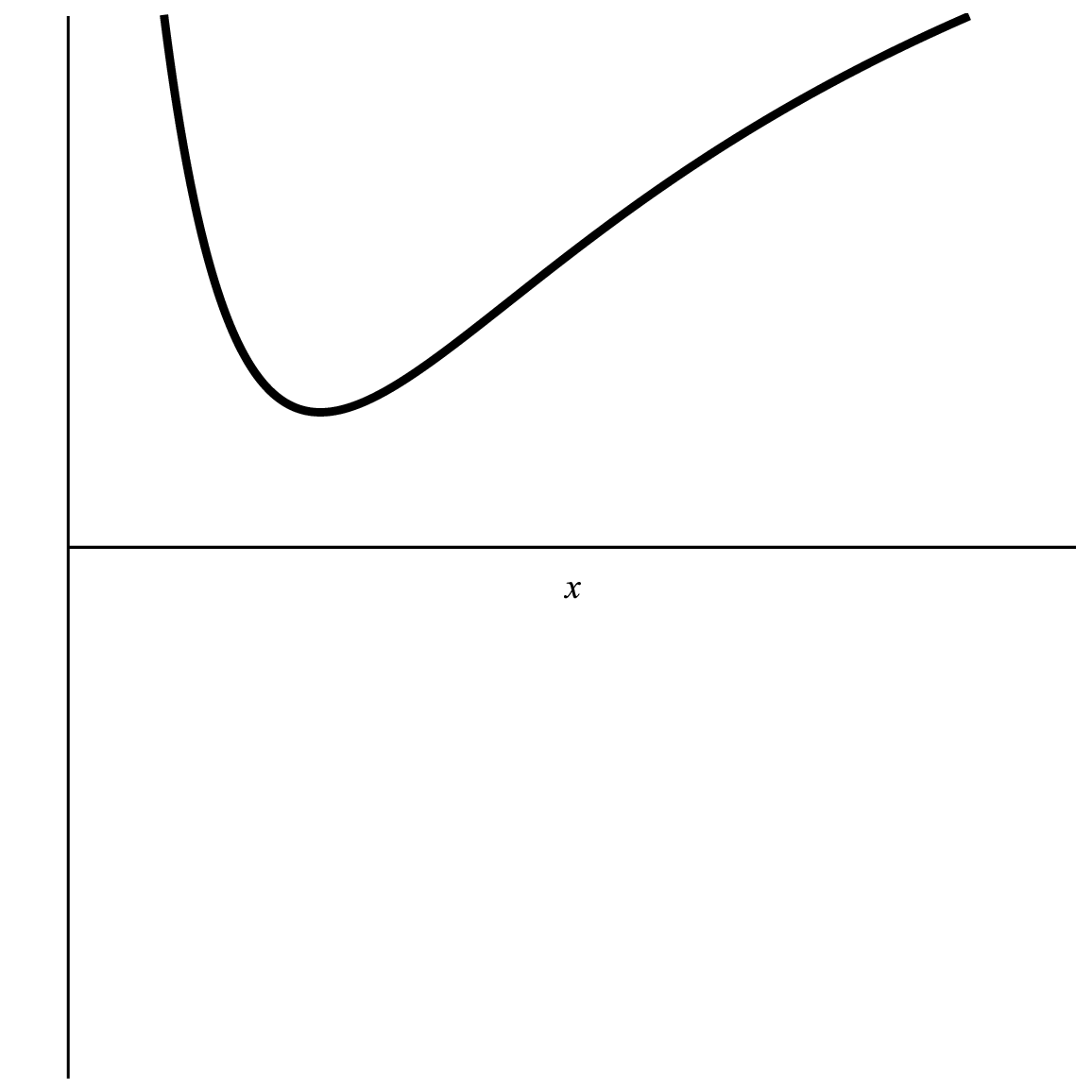}}
\caption{The metric function in Configuration 3, where $m_{\min}<m<m_{c}$
and $q^{2}$ increases from (a) to (f), with $q^{2}=\frac{3}{2}m$ in (e).
Panel (a) admits a simple event horizon and a spacelike singularity at the
center. Panel (b) admits a simple event horizon and a double inner horizon
with a spacelike singularity. Panel (c) represents a black hole with three
simple horizons and a spacelike singularity. Panel (d) admits a double event
horizon and a simple inner horizon with a spacelike singularity. Finally,
panels (e) and (f) represent naked timelike singular solutions with a timelike
singularity at the center.}
\label{F3}
\end{figure}

\begin{figure}[h!]
\centering
\subfloat[]{\includegraphics[width=16cm]{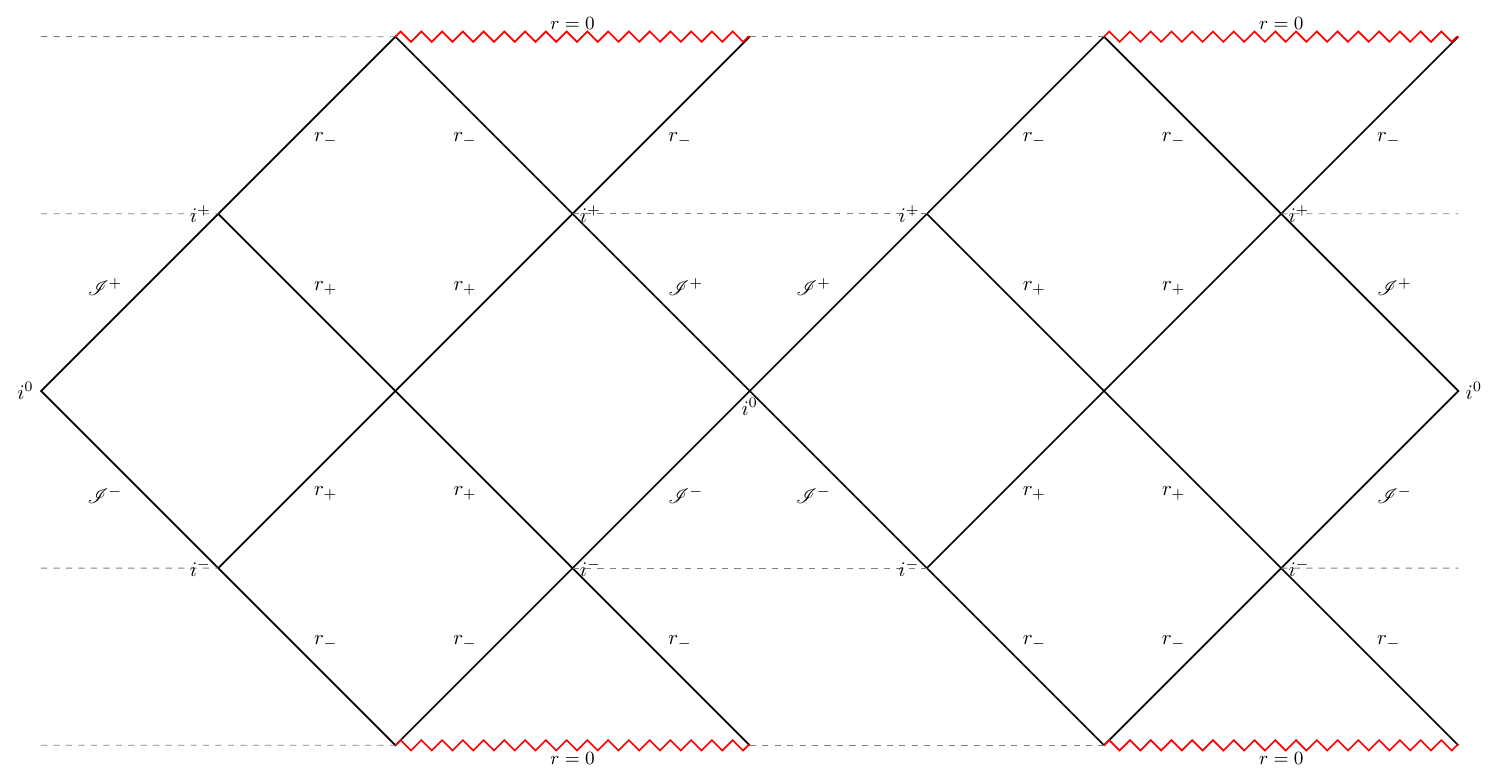}}
\caption{The Penrose diagram of the solutions in Configurations 3 and 4,
corresponding to Fig.~\protect\ref{F3}(b) and Fig.~\protect\ref{F4}(b). The
singularity is spacelike, and $r_{-}$ corresponds to a double horizon.}
\label{P2}
\end{figure}

\begin{figure}[h!]
\centering
\subfloat[]{\includegraphics[width=10cm]{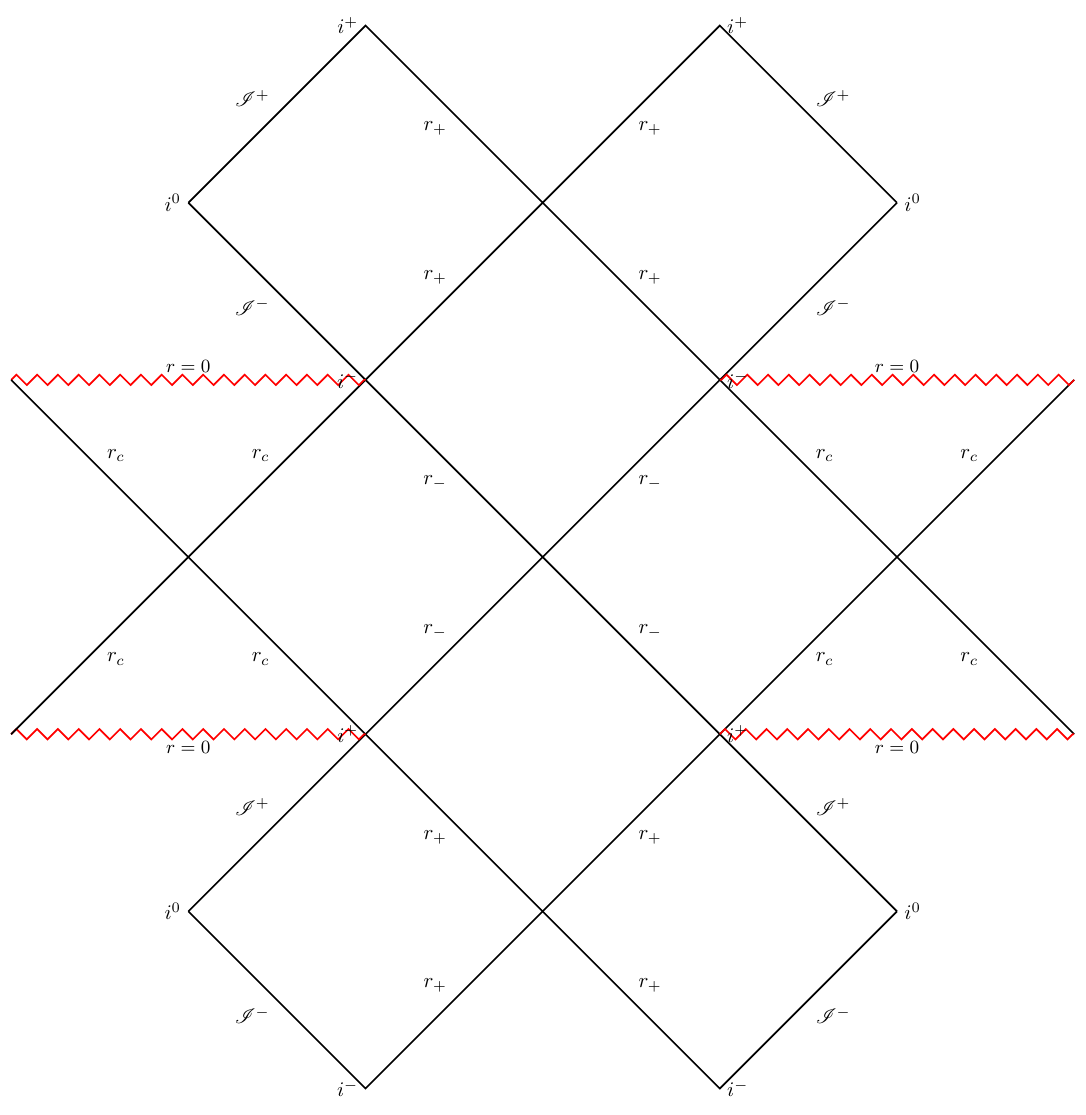}}
\caption{The Penrose diagram of the black hole solutions presented in Fig.~%
\protect\ref{F3}(c) and Fig.~\protect\ref{F4}(c).}
\label{P3}
\end{figure}

\begin{figure}[h!]
\centering
\subfloat[]{\includegraphics[width=10cm]{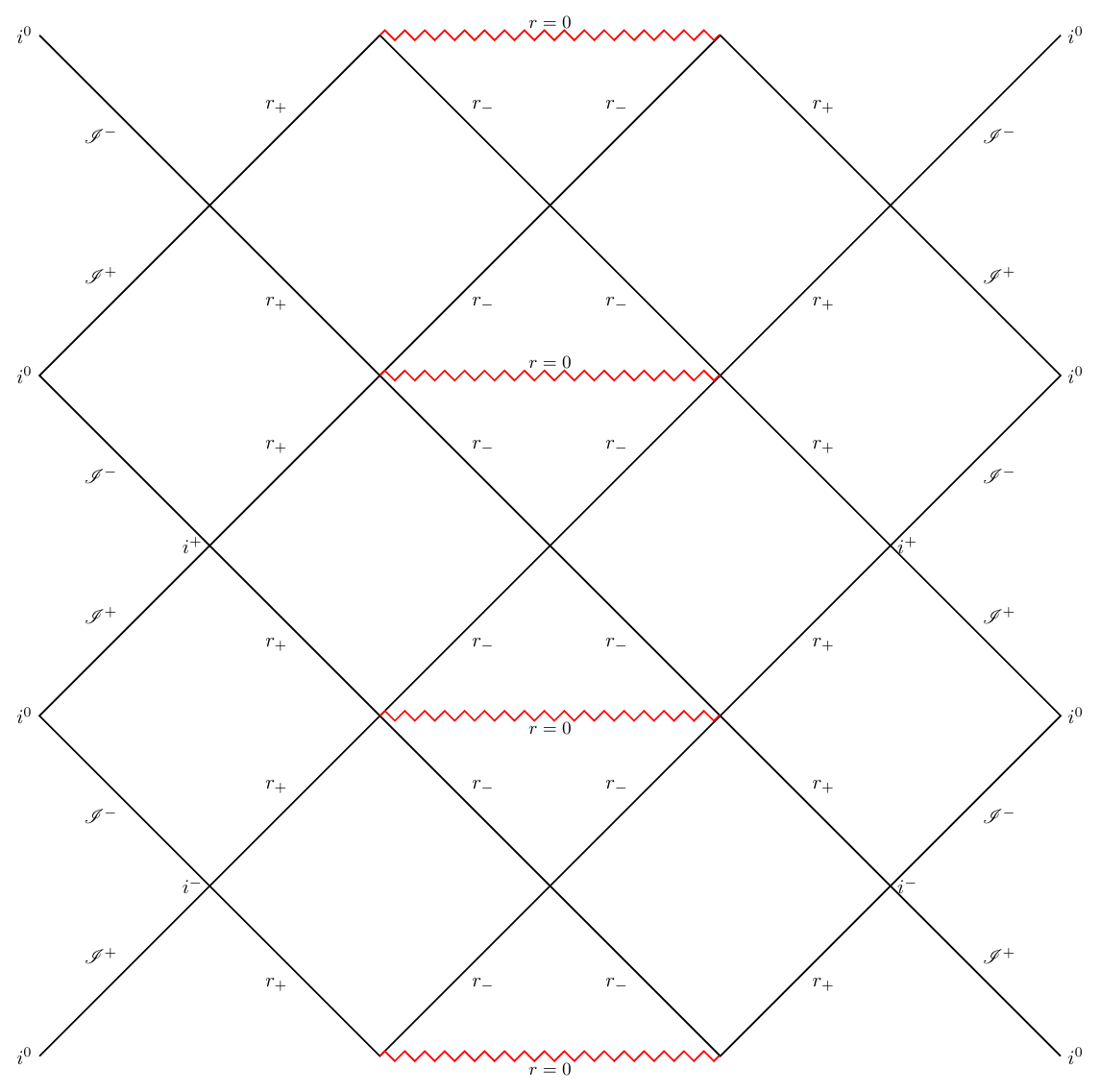}}
\caption{The Penrose diagram of the black hole solutions presented in Fig.~%
\protect\ref{F3}(d).}
\label{P4}
\end{figure}

In the third configuration, $m_{\min }<m<m_{c}$ where there are six
different cases depicted in Fig.~\ref{F3} from (a) to (f): (a) represents a
singular black hole with a single simple event horizon; (b) a black hole
with a simple event horizon and a double inner horizon; (c) a black hole
with three simple horizons (one event horizon and two inner/Cauchy
horizons); (d) a black hole with a double event horizon and a simple inner
horizon; and (e) and (f) represent naked singular solutions. The nature of
the horizons and singularities is illustrated in Figs.~\ref{P1} and \ref{P2}%
. In particular, panel (a) corresponds to Fig.~\ref{P1}(a), panel (b) to
Fig.~\ref{P2}, panel (c) to Fig.~\ref{P3}, panel (d) to Fig.~\ref{P4}, and
panels (e) and (f) to Fig.~\ref{P1}(b).

\subsection{ $m_{c}<m$,}

\begin{figure}[h!]
\centering
\subfloat[]{\includegraphics[width=2.4cm]{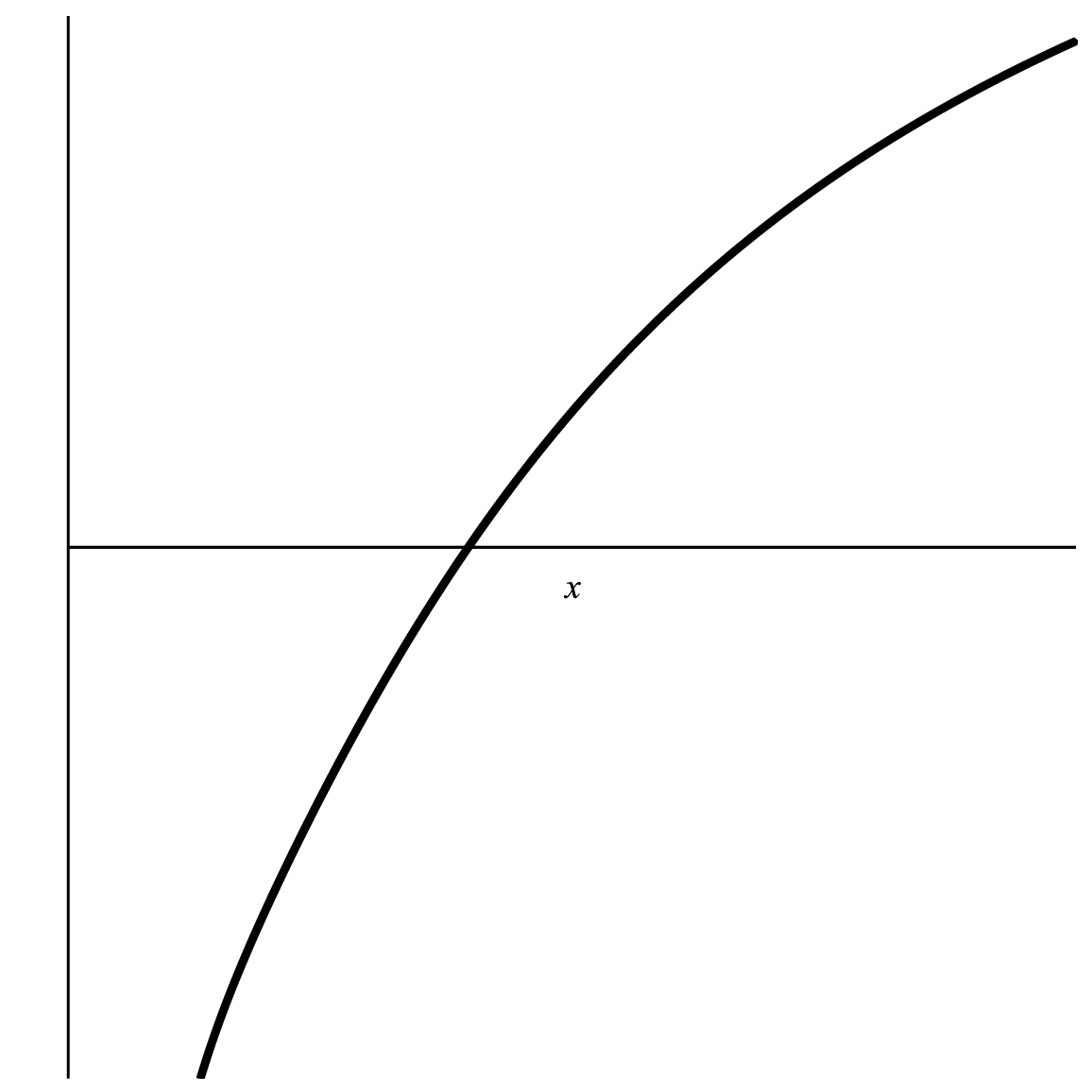}} \qquad %
\subfloat[]{\includegraphics[width=2.4cm]{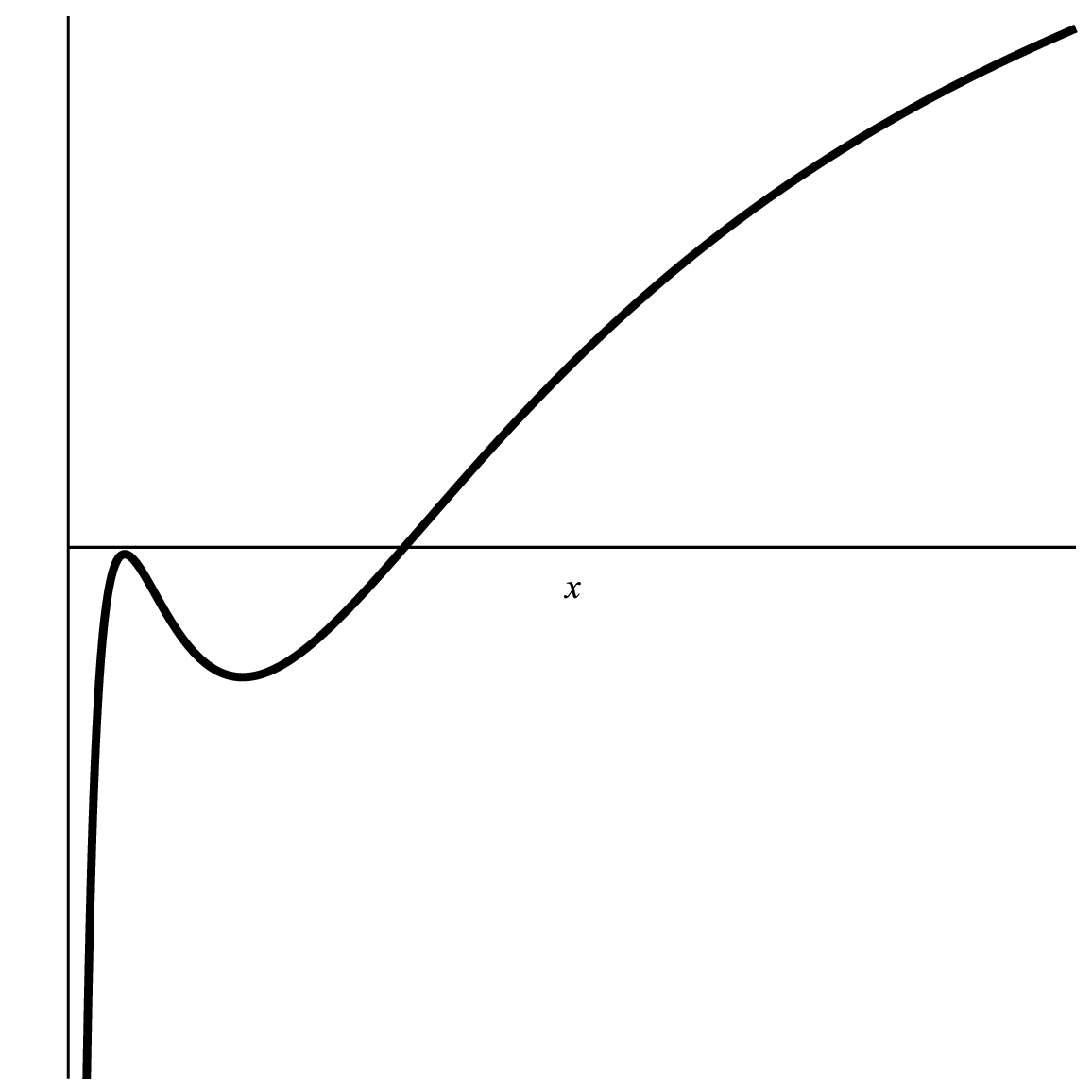}} \qquad %
\subfloat[]{\includegraphics[width=2.4cm]{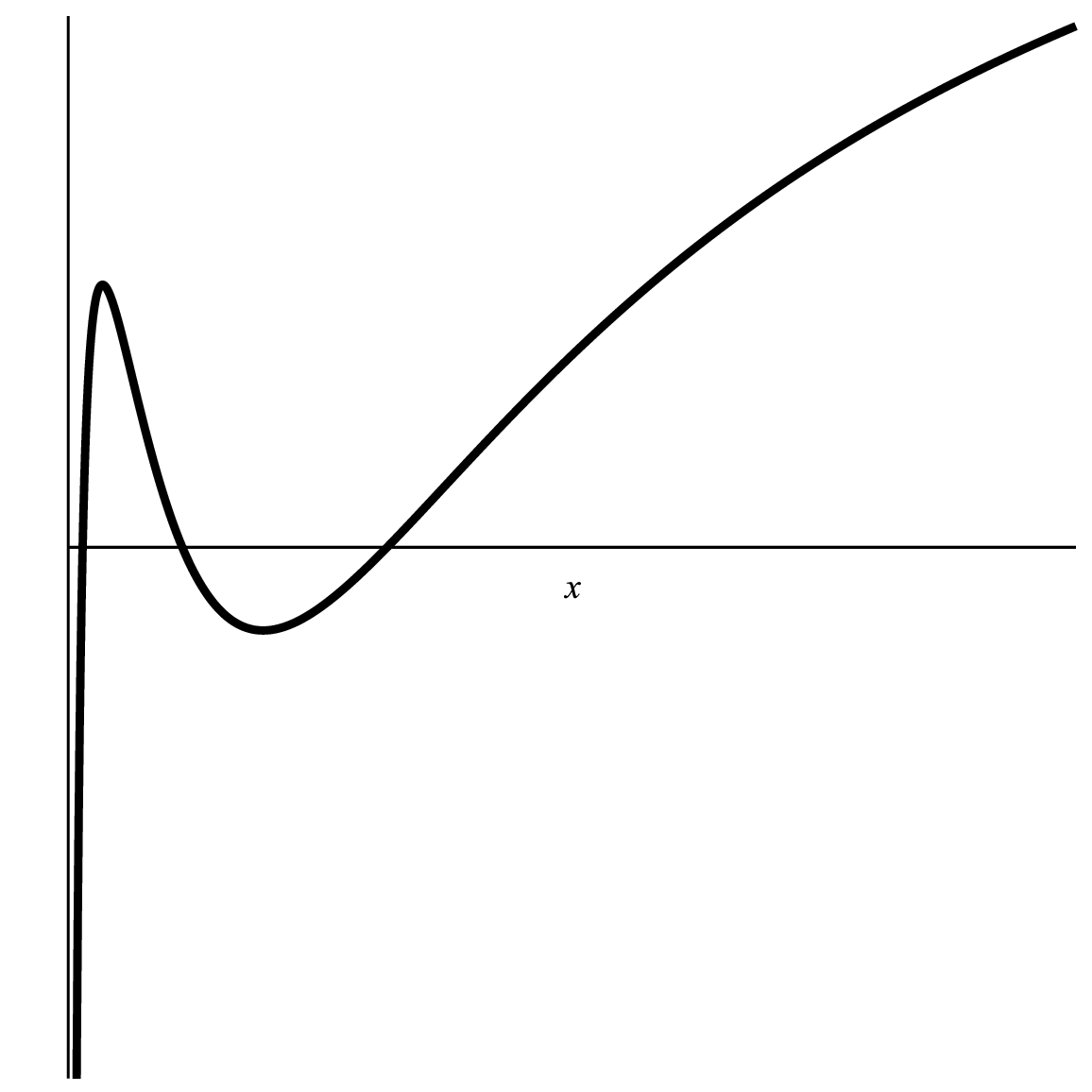}} \qquad %
\subfloat[]{\includegraphics[width=2.4cm]{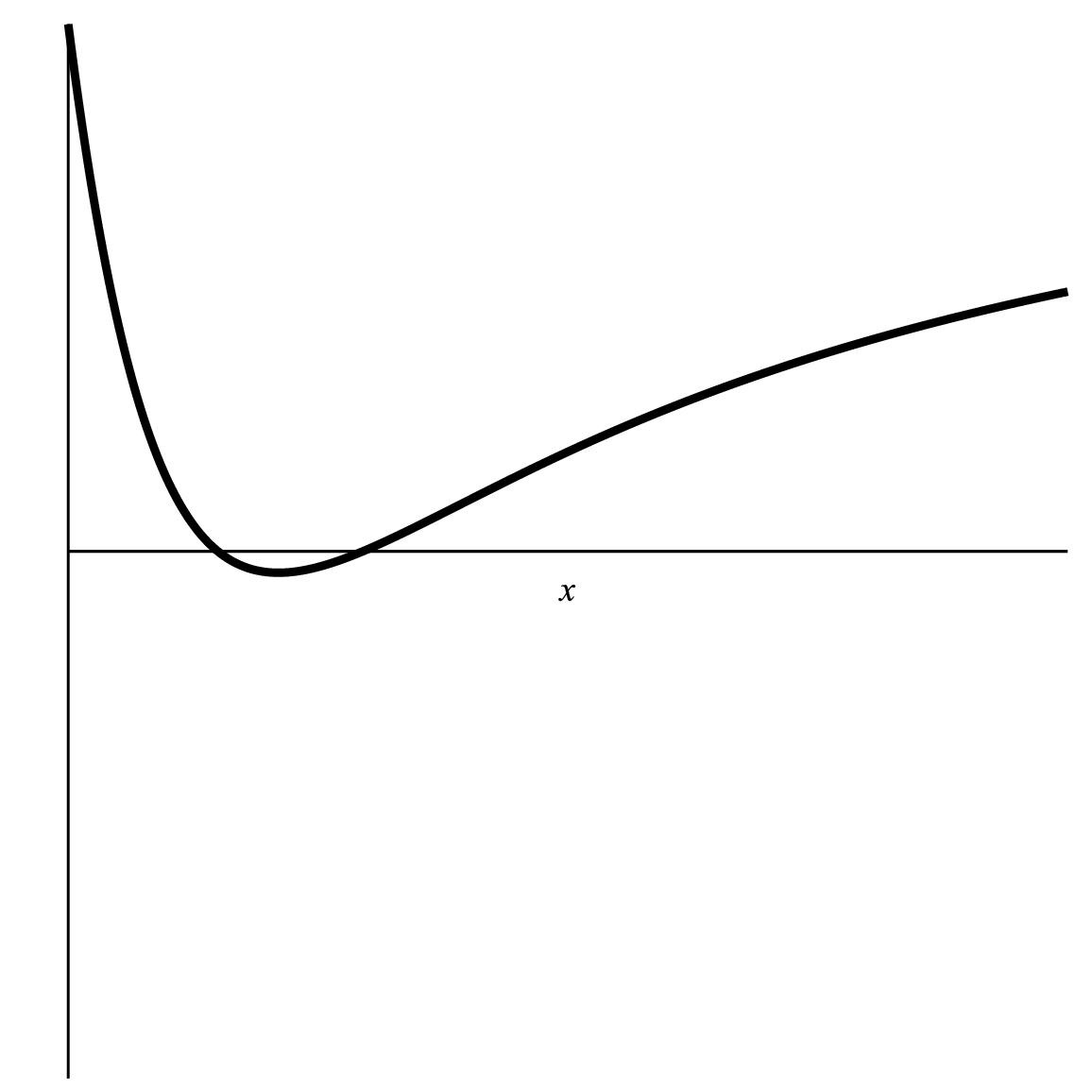}} \qquad %
\subfloat[]{\includegraphics[width=2.4cm]{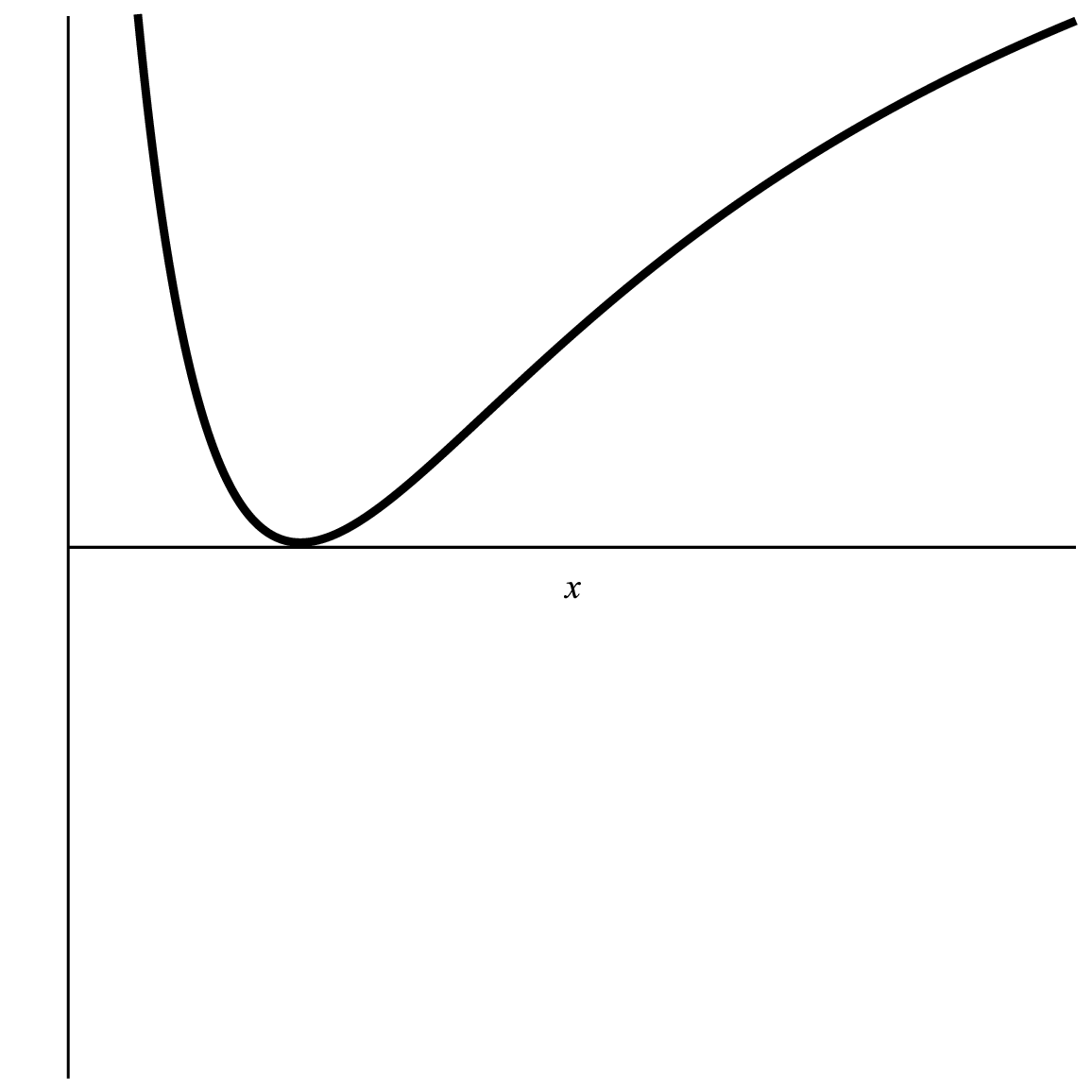}} \qquad %
\subfloat[]{\includegraphics[width=2.4cm]{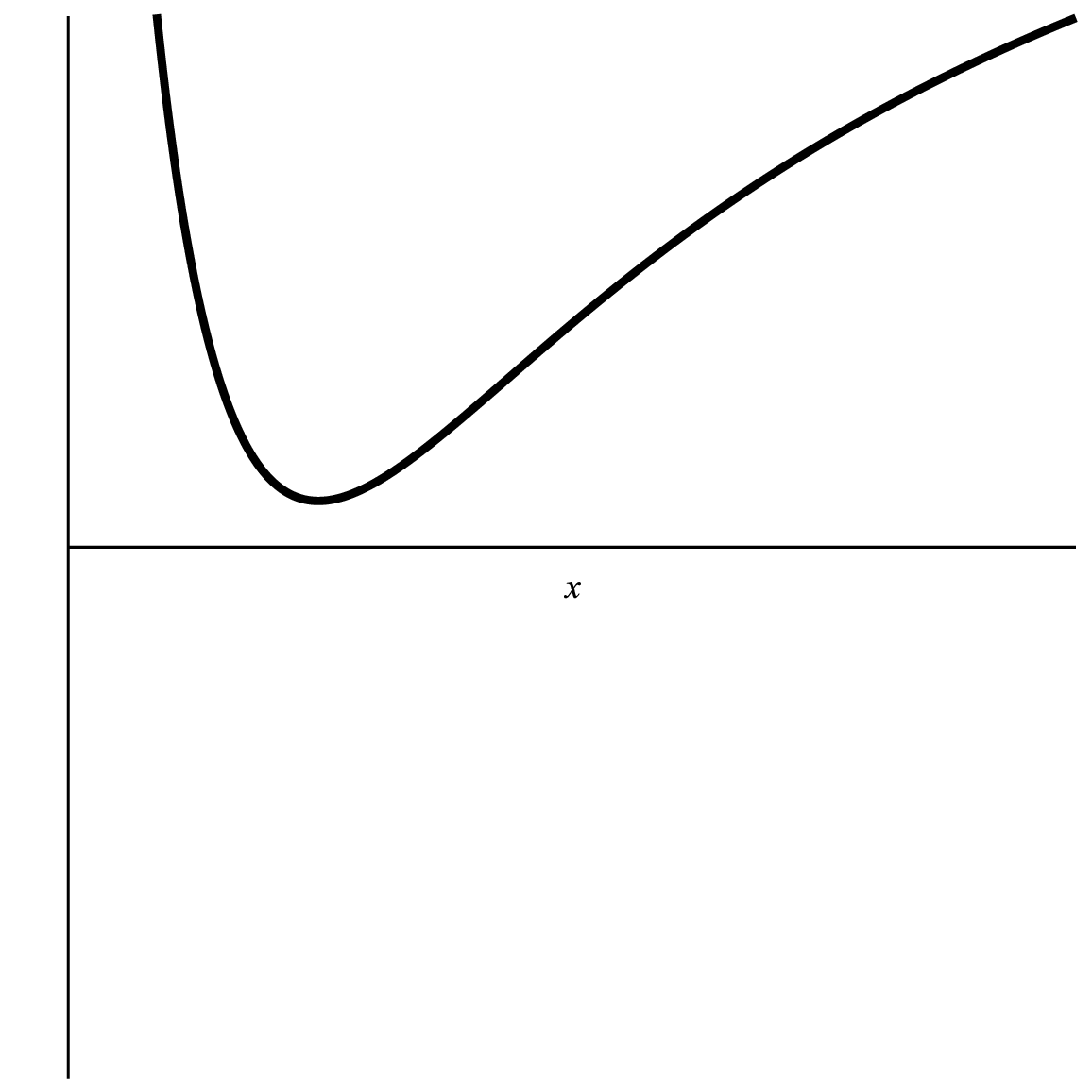}}
\caption{The metric function in Configuration 4, where $m_{c}<m$ and $q^{2}$
increases from (a) to (f), with $q^{2}=\frac{3}{2}m$ in (d). Panels (a),
(b), and (c) are similar to those in Fig.~\protect\ref{F3}. Panel (d) shows
a black hole with two simple horizons and a timelike singularity. Panel (e)
represents an extremal black hole with a double event horizon and a timelike
singularity. Panel (f) corresponds to a naked singular solution with a
timelike singularity at the center.}
\label{F4}
\end{figure}

\begin{figure}[h!]
\centering
\subfloat[]{\includegraphics[width=5cm]{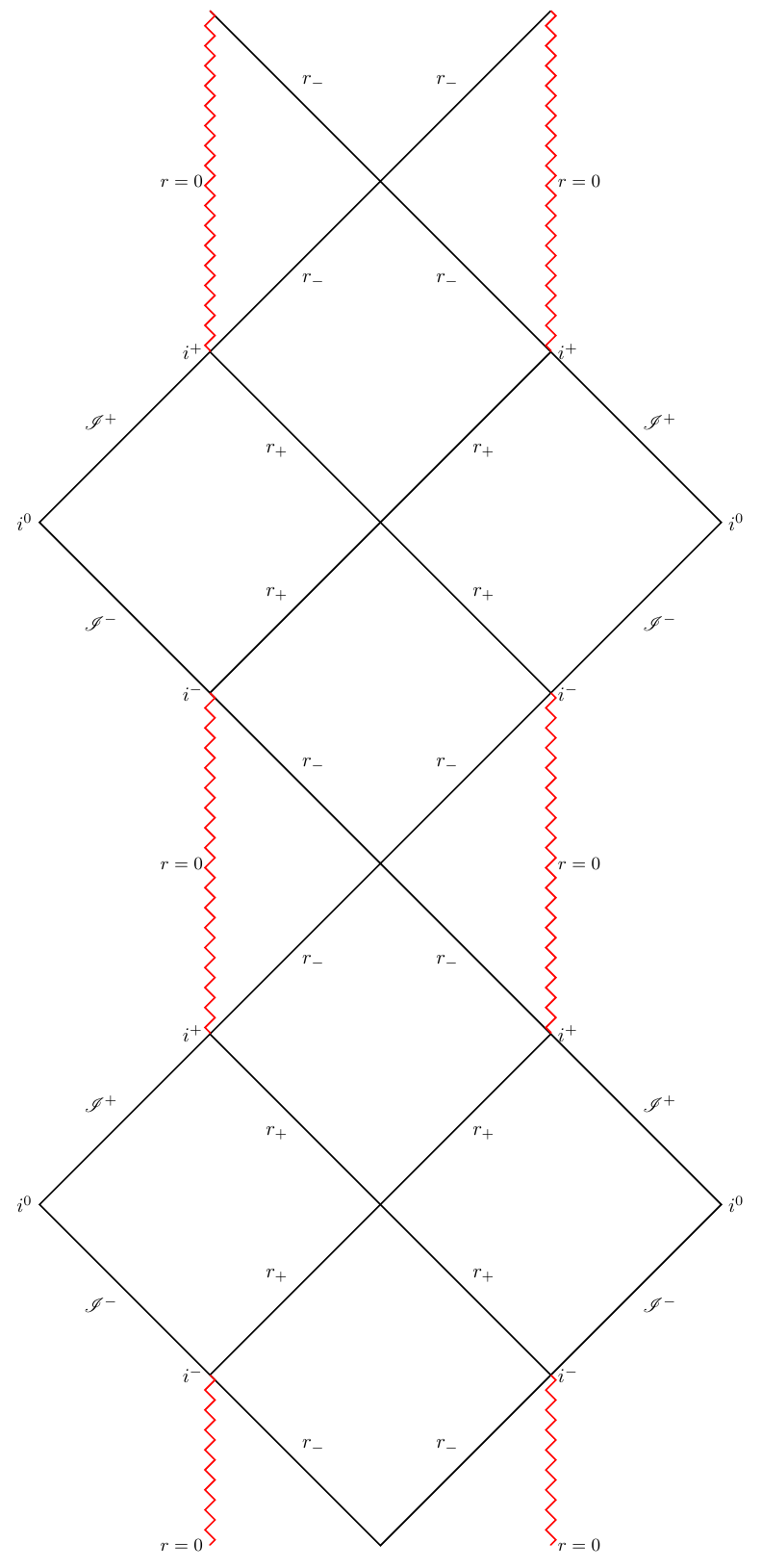}} \qquad %
\subfloat[]{\includegraphics[width=4cm]{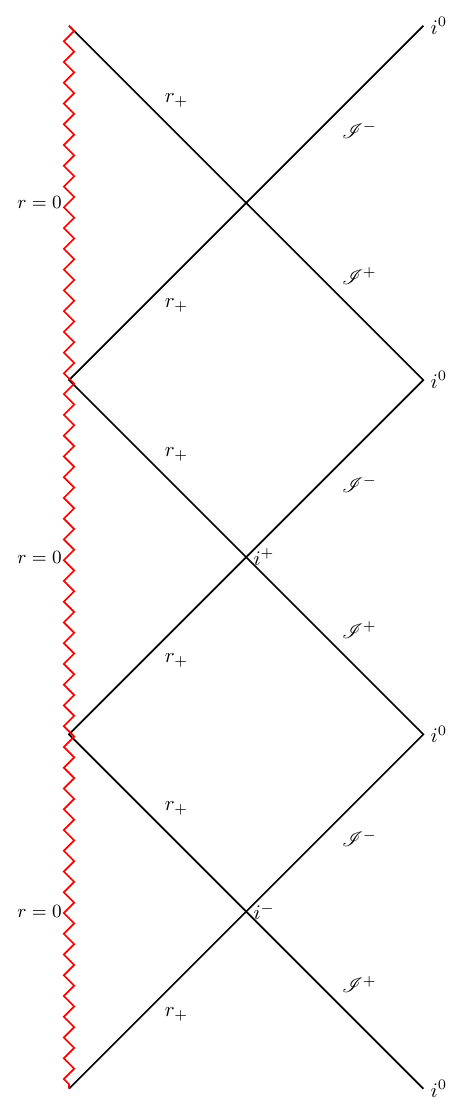}}
\caption{The Penrose diagrams of the black hole solutions presented in Fig.~%
\protect\ref{F4}(d) and (e), corresponding to panels (a) and (b),
respectively.}
\label{P5}
\end{figure}

In the fourth configuration, $m_{c}<m$, as depicted in Fig.~\ref{F4}(a)--(f)
with increasing $q^{2}$. Panels (a)--(e) correspond to black hole solutions,
while panel (f) represents a naked singularity. The black hole solutions in
Fig.~\ref{F4}(a), (b), and (c) are similar to those in Fig.~\ref{F3}(a),
(b), and (c). However, Fig.~\ref{F4}(d) corresponds to $q^{2}=\frac{3m}{2}$
and represents a black hole with two simple horizons (one event horizon and
one Cauchy horizon) and a timelike singularity at the center. The
corresponding Penrose diagram is shown in Fig.~\ref{P5}(a). Another distinct
case is the extremal black hole shown in Fig.~\ref{F4}(e), which admits a
double event horizon and a timelike singularity at the center, as
illustrated in Fig.~\ref{P5}(b). The last panel in Fig.~\ref{F4} is similar
to Fig.~\ref{F3}(f).

\section{Thermal Stability of the Black Hole Solutions}

In this section, we focus on the thermal stability of the black hole
solutions reviewed in the previous section. To this end, we solve $f(r)=0$
to express the ADM mass in terms of the event horizon radius. Referring to
the original metric function (\ref{MF}), we obtain 
\begin{equation}
m=\frac{1}{2}x_{+}+\frac{4q^{2}}{3}\left( x_{+}^{3}-\left( x_{+}^{2}-\frac{1%
}{2}\right) \sqrt{1+x_{+}^{2}}\right) ,
\end{equation}%
where we have used the dimensionless parameters $m=\frac{M}{r_{0}}$, $q^{2}=%
\frac{Q^{2}}{r_{0}^{2}}$, and $x_{+}=\frac{r_{+}}{r_{0}}$. The Hawking
temperature, defined as 
\begin{equation}
T_{H}=\frac{f^{\prime }(r_{+})}{4\pi },
\end{equation}%
is calculated to be 
\begin{equation}
T_{H}=\frac{1}{\pi r_{0}}\left( \frac{1}{4x_{+}}+2q^{2}x_{+}-\frac{%
(1+2x_{+}^{2})q^{2}}{\sqrt{1+x_{+}^{2}}}\right) .
\end{equation}%
Having obtained the Hawking temperature, we compute the heat capacity,
defined by 
\begin{equation}
C=T_{H}\frac{\partial S}{\partial T_{H}},
\end{equation}%
where $S$ is the black hole entropy. In Einstein gravity, minimally coupled
to nonlinear electrodynamics (NED), the black hole entropy is determined
entirely by the gravitational sector and therefore satisfies the standard
Bekenstein-Hawking area law~\cite{Bekenstein1973,Hawking1975}, 
\begin{equation}
S = \frac{A}{4} = \pi r_{+}^{2}.
\end{equation}
This result follows rigorously from Wald's Noether charge formalism~\cite%
{Wald1993,IyerWald1994}, which shows that the entropy depends only on the
variation of the gravitational Lagrangian with respect to the Riemann
tensor. Since the NED contribution enters solely through the matter sector
and does not involve curvature couplings, it does not modify the entropy.
Consequently, while nonlinear electrodynamics can significantly affect the
spacetime geometry, horizon structure, and thermodynamic quantities such as
temperature and heat capacity, the entropy retains its universal area-law
form in Einstein--NED theories~\cite%
{Rasheed1997,Breton2005,Dey2004,Hendi2012}. Hence, we obtain 
\begin{equation}
C = \frac{2\pi r_{0}^{2}x_{+}^{2}(1+x_{+}^{2}) \left[ \sqrt{1+x_{+}^{2}} +
4q^{2}x_{+}\left( 2x_{+}\left( \sqrt{1+x_{+}^{2}} - x_{+}\right) - 1\right) %
\right]}{4x_{+}^{2}q^{2}\left[ 2(1+x_{+}^{2})^{3/2} - x_{+}(2x_{+}^{2}+3) %
\right] - (1+x_{+}^{2})^{3/2}}.
\end{equation}

\begin{figure}[h]
\centering
\subfloat[]{\includegraphics[width=5cm]{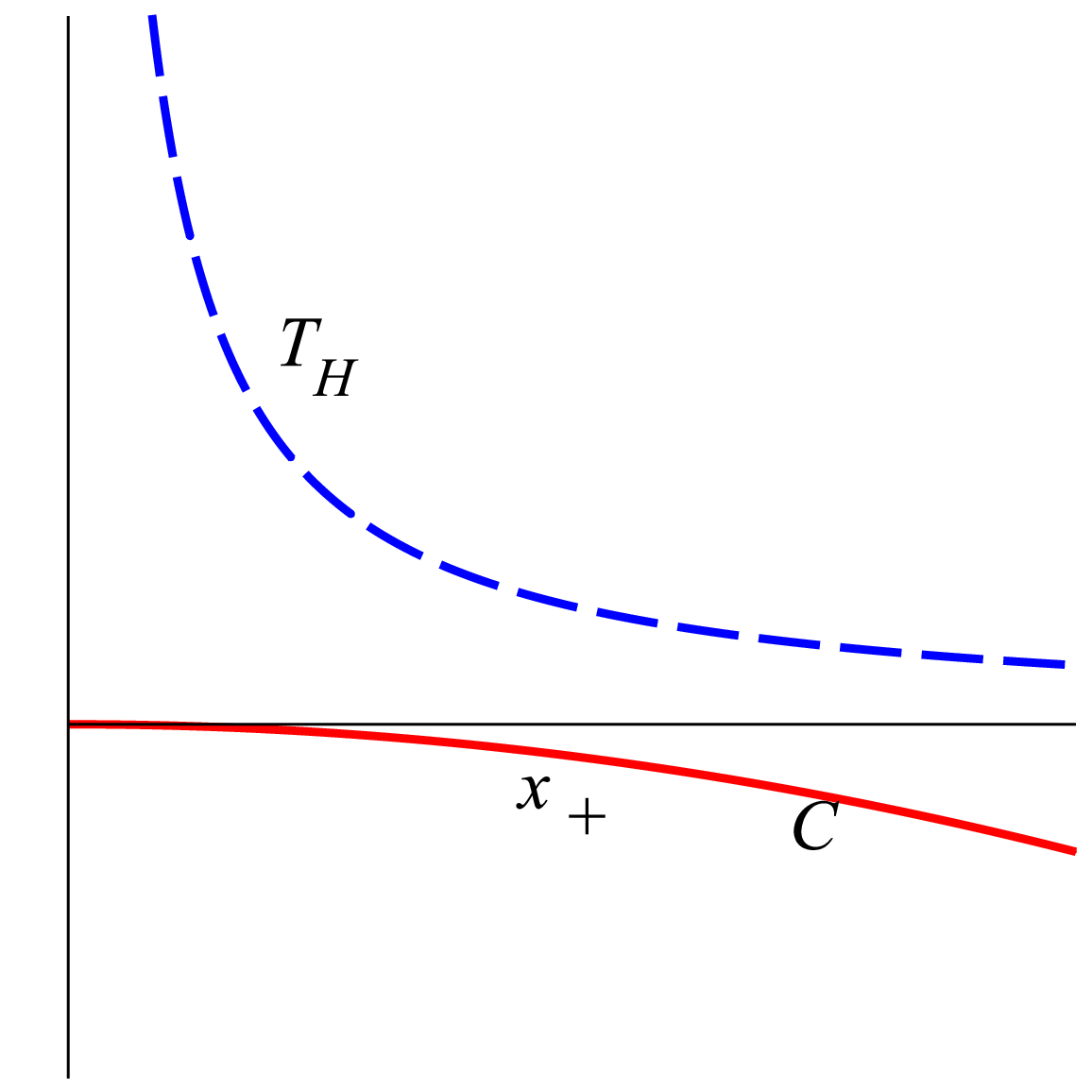}} \qquad \subfloat[]{%
\includegraphics[width=5cm]{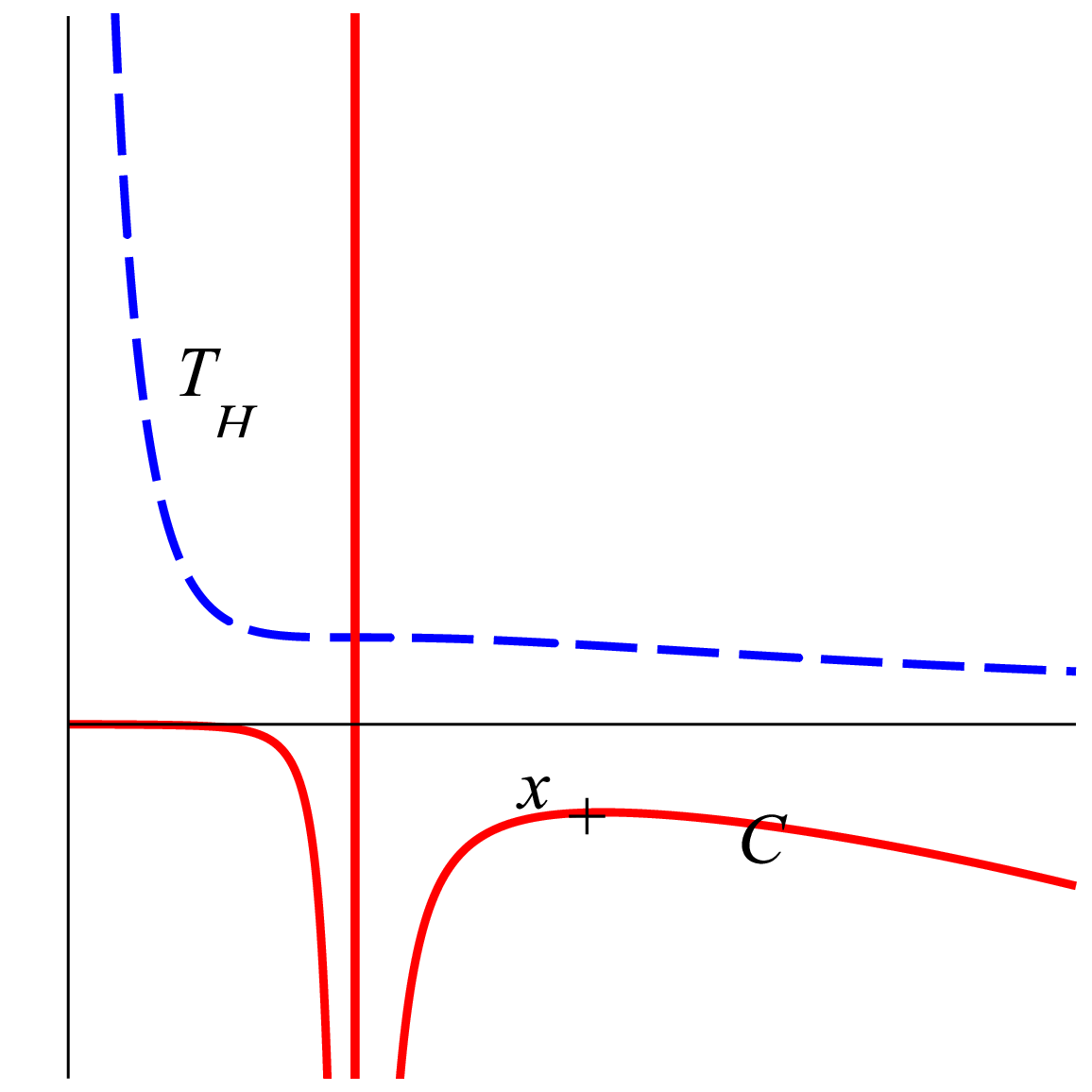}} \qquad \subfloat[]{%
\includegraphics[width=5cm]{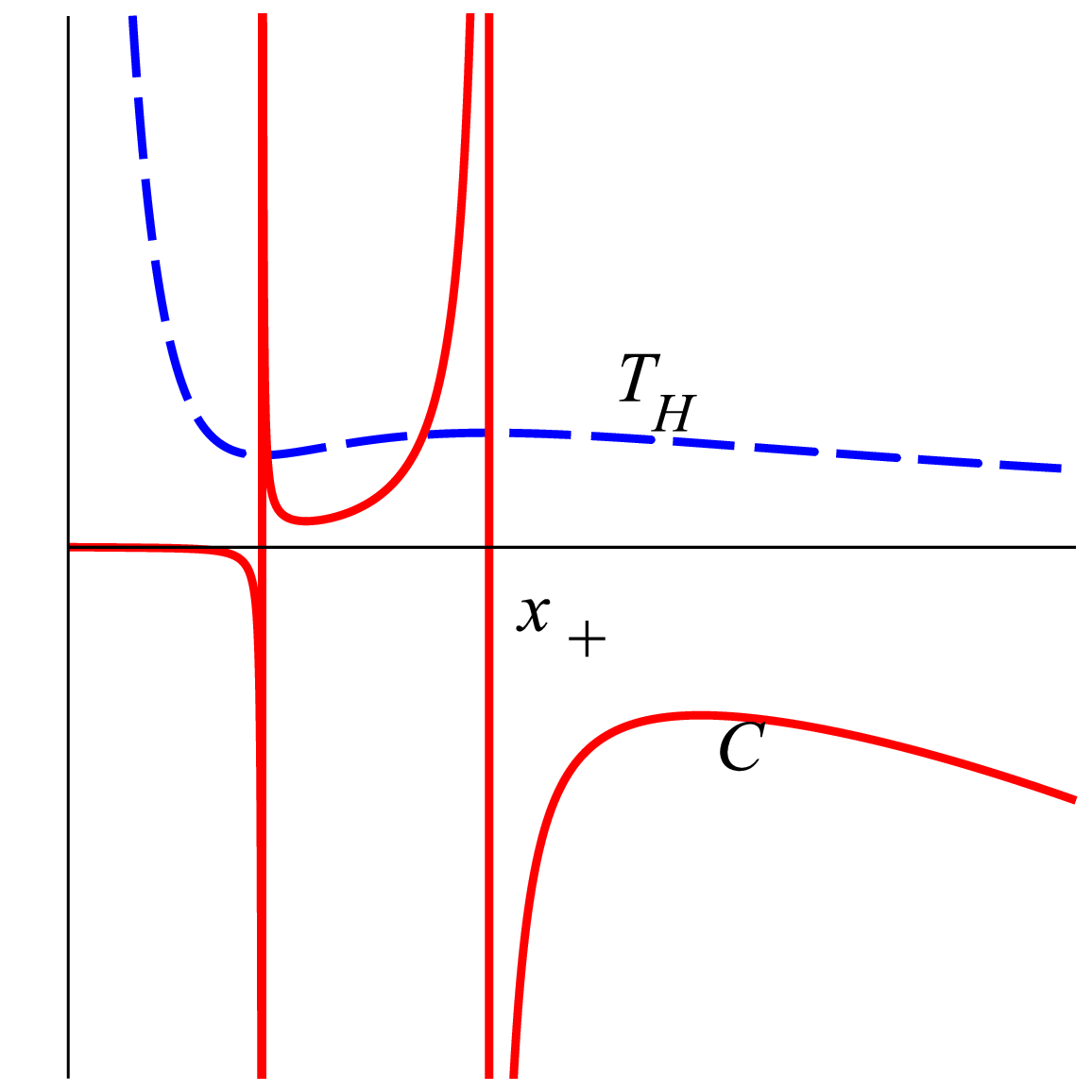}} \qquad \subfloat[]{%
\includegraphics[width=5cm]{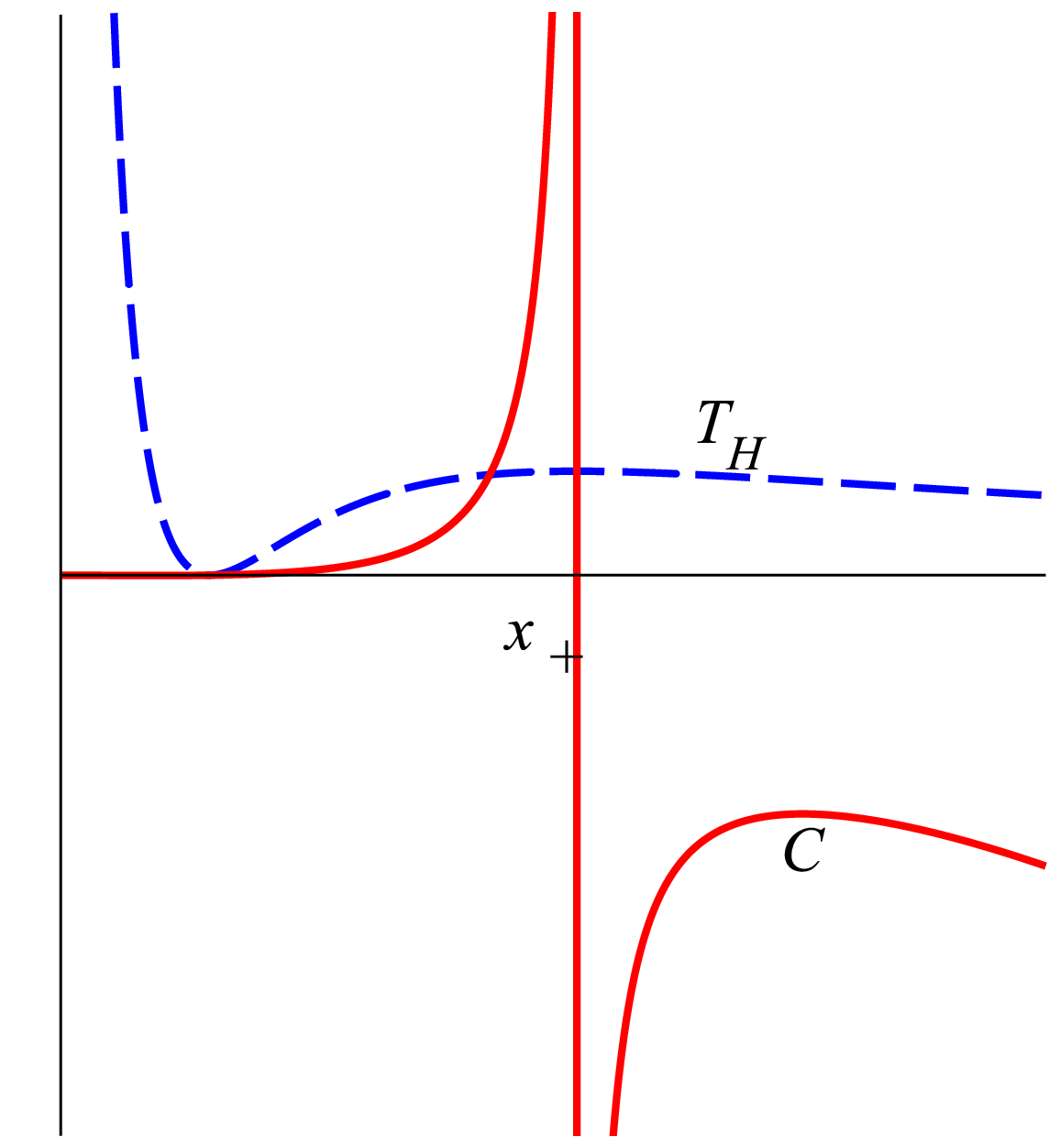}} \qquad \subfloat[]{%
\includegraphics[width=5cm]{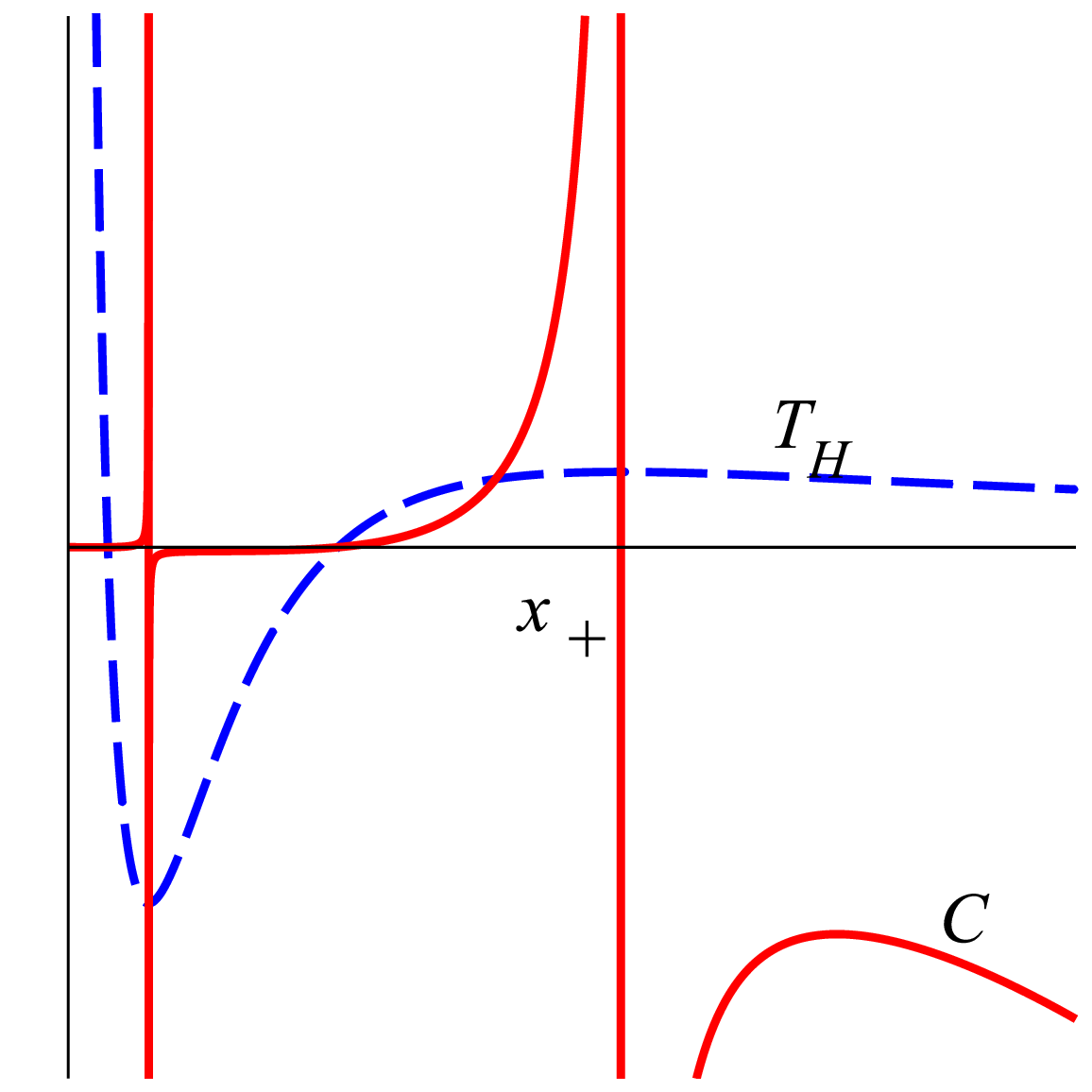}}
\caption{Generic plots of the Hawking temperature and heat capacity versus $%
x_{+}$, with $q^{2}=0$ in panel (a) and increasing up to panel (e). As the
electric charge increases and becomes greater than $q_{c}$, an initially
unstable ( Panels (a) and (b) ) black hole develops a thermally stable
region between the two transition points of the heat capacity (Panel (c)).
For larger values of $q^{2}$ such that $q^{2}\geq q_{\min }^{2}$, the
mimimum of Hawking temperature reaches zero or negative values, and the
region of thermal stability becomes confined between the larger zero point
of the Hawking temperature and the larger transition point (Panels (d) and
(e)).}
\label{F10}
\end{figure}

Next, we analyze the thermal stability of the black hole by examining the
signs of both the heat capacity and the Hawking temperature. The black hole
is thermally stable when both $C$ and $T_{H}$ are positive. In Fig.~\ref{F10}%
, we plot the Hawking temperature and heat capacity as functions of $x_{+}$
for increasing values of $q^{2}$ from (a) to (e). As shown in panels (a) and
(b), when $q^{2}$ increases from zero in (a) up to a critical value in (c)
where $q^{2}=q_{c}^{2}=1.052647$, the Hawking temperature is a decreasing
function of $x_{+}$, while the heat capacity remains negative and
decreasing. Therefore, below the critical value of $q^{2}$, the black hole
is thermally unstable. This behavior changes once $q^{2}$ exceeds the
critical value i.e., $q_{c}^{2}$. In this regime, the Hawking temperature
develops two extrema, namely a local minimum followed by a local maximum. At
these points, the heat capacity diverges and changes sign, indicating phase
transitions. In particular, the heat capacity becomes positive between these
two transition points, signaling a thermally stable phase. For values of $%
q^{2}$ greater than a second critical value $q_{\min }^{2}=\frac{3}{4}+\frac{%
\sqrt{2}}{2}$, the Hawking temperature remains positive only up to a certain
range, beyond which its minimum becomes negative. This indicates that the
second critical value $q_{\min }^{2}$ corresponds to the formation of a
degenerate (extremal) horizon. This is the minimum electric charge upon
which an extremal black hole may form. Consequently, the black hole is
thermally stable only for horizon radii lying between this extremal point
and the larger transition point of the heat capacity. This behavior is
illustrated in panels (d) and (e) of Fig.~\ref{F10}.

Therefore, we observe that a black hole solution is thermally stable if $%
q^{2}>q_{c}^{2}$ within a finite interval of the event horizon radius. This
interval lies either between the two extrema of the Hawking temperature -
known as Davies points - at which the heat capacity diverges, or between the
radius of the extremal black hole (where the Hawking temperature vanishes)
and the larger Davies point.

Before concluding this section, we examine the relation between the ADM mass
and the electric charge for the existence of black hole solutions in the
general case. To this end, we return to the original form of the metric
function~(\ref{MF}) and introduce the dimensionless quantities $Q = M\alpha$%
, $r = M\eta$, and $r_{0} = M\eta_{0}$. This yields 
\begin{equation}
f(\eta) = 1 - \frac{2}{\eta} + \frac{8\alpha^{2}}{3\eta_{0}^{2}} \left( 
\frac{\eta^{2}}{\eta_{0}^{2}} - \frac{\frac{\eta^{2}}{\eta_{0}^{2}} - \frac{1%
}{2}}{\frac{\eta}{\eta_{0}}} \sqrt{1 + \frac{\eta^{2}}{\eta_{0}^{2}}}
\right).
\end{equation}

Solving $f(\eta_{+}) = 0$ for $\alpha$, with $\eta = \eta_{+} = \frac{r_{+}}{%
M}$, we obtain 
\begin{equation}
\alpha = \pm \eta_{0}^{2} \sqrt{\frac{\eta_{+} - 2}{\left( 2\eta_{+}^{2} -
\eta_{0}^{2} \right) \sqrt{\eta_{+}^{2} + \eta_{0}^{2}} - 2\eta_{+}^{3}}}.
\end{equation}

\begin{figure}[h]
\centering
\subfloat[]{\includegraphics[width=8cm]{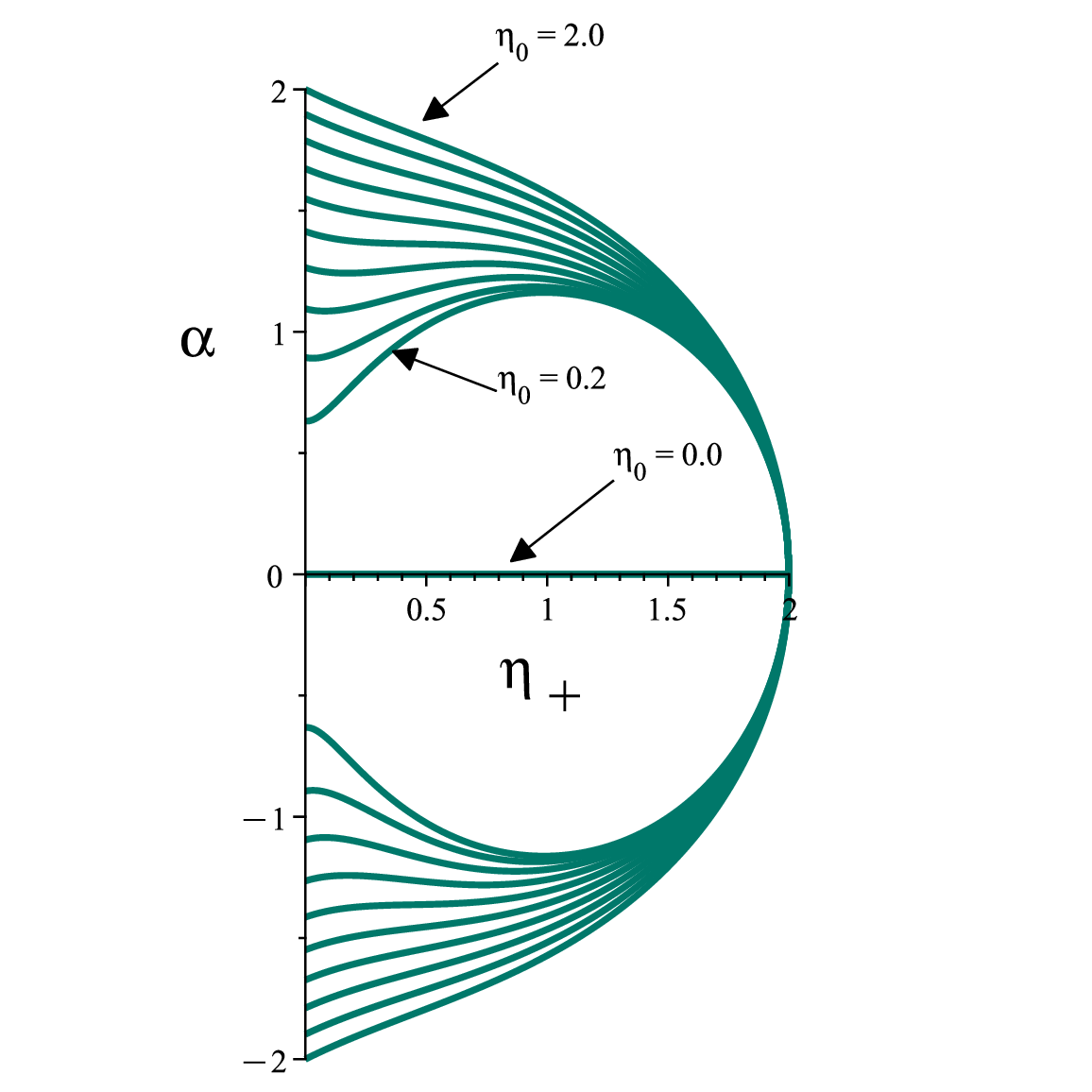}} \qquad %
\subfloat[]{\includegraphics[width=8cm]{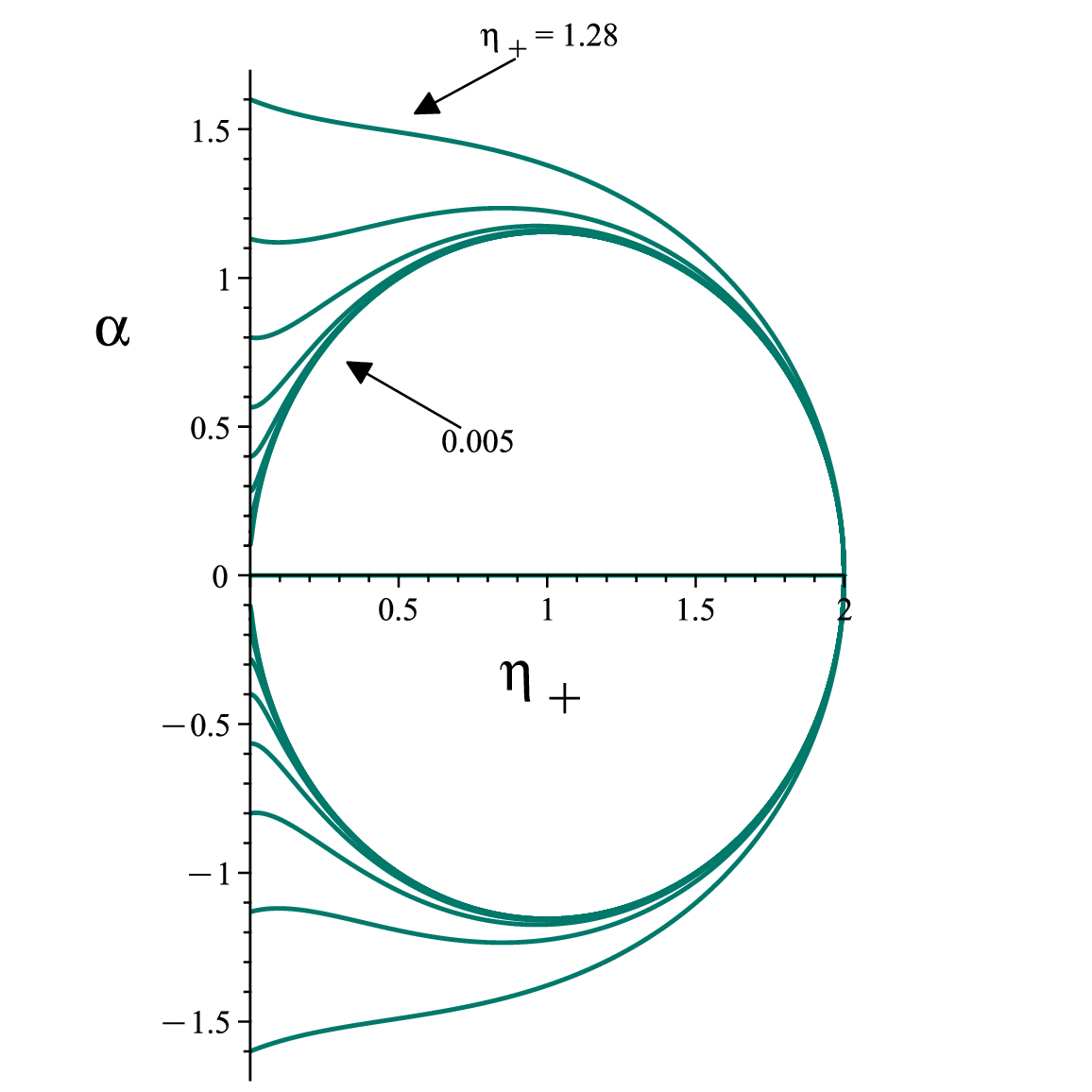}}
\caption{Plots of $\protect\alpha = \frac{Q}{M}$ versus $\protect\eta_{+} = 
\frac{r_{+}}{M}$ for different values of $\protect\eta_{0} = \frac{r_{0}}{M}$%
. Panel (a) shows $\protect\eta_{0}$ varying from $0.0$ to $2.0$ with equal
intervals, while panel (b) corresponds to $\protect\eta_{0} = 0.005, 0.01,
0.02, 0.04, 0.08, 0.16, 0.32, 0.64,$ and $1.28$.}
\label{F11}
\end{figure}

In Fig.~\ref{F11}, we plot $\alpha = \frac{Q}{M}$ as a function of $\eta_{+}
= \frac{r_{+}}{M}$ for various values of $\eta_{0} = \frac{r_{0}}{M}$. We
observe that, irrespective of the values of $\eta_{0}$ and $\alpha$, the
horizon radius $\eta_{+}$ does not exceed $2$, and approaches $\eta_{+} = 2$
in the limit $\alpha = 0$. On the other hand, as the ratio $\alpha = \frac{Q%
}{M}$ increases, the horizon radius decreases. For each fixed value of $%
\eta_{0}$, there exists a critical value of $\alpha$ beyond which no black
hole solution exists.

Furthermore, for values of $\eta_{0}$ smaller than a certain critical value,
the solution admits multiple horizons within a finite interval of $\alpha$.

\section{Near-Event-Horizon Geometry}

The black hole solutions presented in Figs.~\ref{F1}, \ref{F2}, \ref{F3},
and \ref{F4} admit different types of event horizons: non-degenerate
horizons for which $f(r_{+})=0$, double-degenerate horizons satisfying $%
f(r_{+})=f^{\prime }(r_{+})=0$, and triple-degenerate horizons for which $%
f(r_{+})=f^{\prime }(r_{+})=f^{\prime \prime }(r_{+})=0$. In our solutions,
most configurations possess non-degenerate horizons, except for Fig.~\ref{F3}%
(d) and Fig.~\ref{F4}(e), which correspond to double-degenerate (extremal)
black holes, and Fig.~\ref{F2}(b), which represents a triple-degenerate
(ultra-extremal) black hole. Recalling that the Ricci scalar and the
Kretschmann scalar associated with the metric~(\ref{g}) are given by 
\begin{equation}
\mathcal{R}=-f^{\prime \prime }-\frac{4}{r}f^{\prime }-\frac{2f}{r^{2}}+%
\frac{2}{r^{2}},
\end{equation}%
and 
\begin{equation}
\mathcal{K}=\left( f^{\prime \prime }\right) ^{2}+\frac{4}{r^{2}}\left(
f^{\prime }\right) ^{2}+\frac{4f^{2}}{r^{4}}-\frac{8f}{r^{4}}+\frac{4}{r^{4}}%
,
\end{equation}%
we find that, at the double-degenerate horizon, 
\begin{equation}
\mathcal{R}=-f^{\prime \prime }(r_{+})+\frac{2}{r_{+}^{2}},\qquad \mathcal{K}%
=\left( f^{\prime \prime }(r_{+})\right) ^{2}+\frac{4}{r_{+}^{4}},
\end{equation}%
while for the triple-degenerate horizon, 
\begin{equation}
\mathcal{R}=\frac{2}{r_{+}^{2}},\qquad \mathcal{K}=\frac{4}{r_{+}^{4}}=%
\mathcal{R}^{2}.
\end{equation}%
This shows that, at the triple-degenerate horizon, the curvature invariants
take particularly simple forms and are completely determined by the horizon
radius. Our main interest lies in the triple-degenerate (ultra-extremal)
case. Expanding the metric function near $r=r_{+}$, we obtain 
\begin{equation}
f(r)\simeq \frac{1}{6}f^{\prime \prime \prime }(r_{+})(r-r_{+})^{3}.
\end{equation}%
Consequently, the near-horizon geometry is described by 
\begin{equation}
ds^{2}=-\frac{1}{6}f^{\prime \prime \prime }(r_{+})(r-r_{+})^{3}dt^{2}+\frac{%
6}{f^{\prime \prime \prime }(r_{+})(r-r_{+})^{3}}dr^{2}+r_{+}^{2}\left(
d\theta ^{2}+\sin ^{2}\theta \,d\phi ^{2}\right) .
\end{equation}%
The near-horizon geometry of the triple-degenerate horizon has the topology 
\begin{equation}
\mathbb{R}_{t}\times \mathbb{R}_{+}\times S^{2}\simeq \mathbb{R}^{2}\times
S^{2},
\end{equation}%
which is identical to that of standard static, spherically symmetric black
holes. However, in contrast to the extremal case, the $(t,r)$ sector is not
maximally symmetric and does not reduce to $\mathrm{AdS}_{2}$. Instead, it
describes a two-dimensional geometry with non-constant curvature and a
stronger near-horizon scaling of the form $f(r)\sim (r-r_{+})^{3}$. As a
consequence, the proper radial distance to the horizon diverges, indicating
that the horizon lies at an infinite proper distance, similar to extremal
black holes but with a stronger divergence. Moreover, the redshift effect
becomes more pronounced due to the higher-order vanishing of the metric
function. It is worth emphasizing that the near-horizon geometry is
sufficient to determine certain thermodynamic quantities, such as the
entropy and the Hawking temperature. In the present case, the
triple-degenerate horizon satisfies $f^{\prime }(r_{+})=0$, which implies a
vanishing surface gravity and hence zero Hawking temperature, confirming
that the solution represents an ultra-extremal black hole. Finally, in
contrast to the extremal RN black hole, whose near-horizon geometry $\mathrm{%
AdS}_{2}\times S^{2}$ is locally isometric to the interaction region of
certain colliding plane wave solutions (e.g., the Bell-Szekeres spacetime),
the present triple-degenerate configuration does not admit such a
correspondence. This is because the $(t,r)$ sector is not maximally
symmetric and does not reduce to $\mathrm{AdS}_{2}$, thereby preventing any
local mapping to colliding plane wave geometries.

\section{Shadow of the Black Hole}
\begin{figure}[h]
\centering\subfloat[]{\includegraphics[width=7cm]{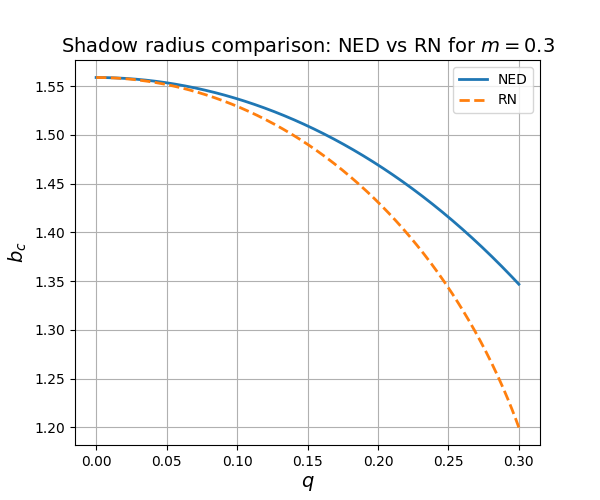}} %
\subfloat[]{\includegraphics[width=7cm]{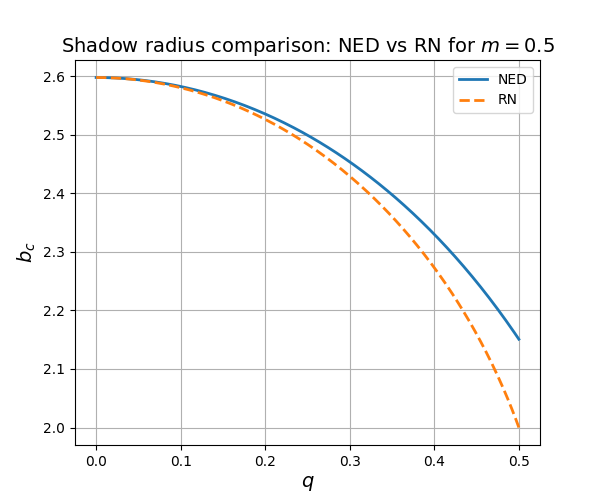}}\qquad\subfloat[]{%
\includegraphics[width=7cm]{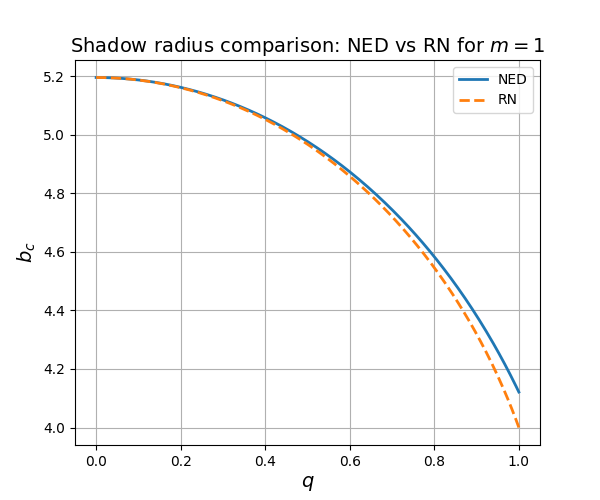}}\subfloat[]{%
\includegraphics[width=7cm]{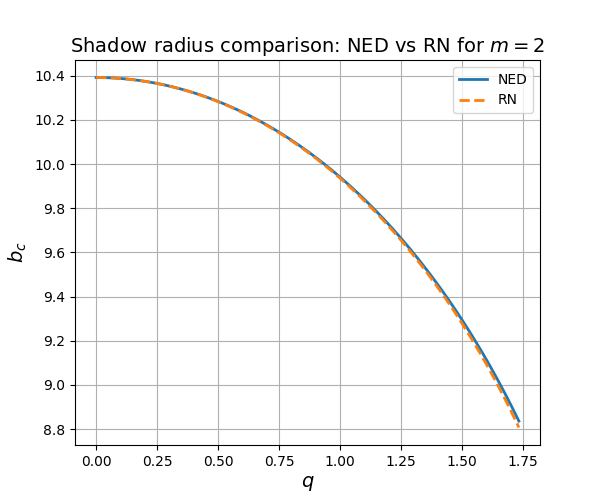}}
\caption{The radius of the shadow observed at infinity in terms of the
dimensionless electric charge $q$, with $m=0.3$ (a)$,$ $m=0.5$ (b), $m=1.0$
(c)$,$ and $m=2.0$ (d). The dashed curve depicts the Reissner-Nordstr\"{o}m
case while the solid curve represents the NED black holes.}
\label{F12}
\end{figure}
The shadow of a black hole is determined by unstable circular null geodesics
(the photon sphere). To derive the corresponding condition, we consider null
geodesics in the static, spherically symmetric spacetime 
\begin{equation}
ds^{2}=-f(r)\,dt^{2}+\frac{1}{f(r)}\,dr^{2}+r^{2}\left( d\theta ^{2}+\sin
^{2}\theta \,d\phi ^{2}\right) .
\end{equation}%
Without loss of generality, we restrict the motion to the equatorial plane $%
\theta =\pi /2$. The Lagrangian for a test particle is then given by 
\begin{equation}
\mathcal{L}=-f(r)\dot{t}^{2}+\frac{\dot{r}^{2}}{f(r)}+r^{2}\dot{\phi}^{2},
\end{equation}%
where a dot denotes differentiation with respect to an affine parameter.
Since the metric is independent of $t$ and $\phi $, there exist two
conserved quantities 
\begin{equation}
E=f(r)\dot{t},\text{ and }L=r^{2}\dot{\phi},
\end{equation}%
which correspond to the energy and angular momentum of the photon,
respectively. For null geodesics ($ds^{2}=0$), we obtain 
\begin{equation}
-f(r)\dot{t}^{2}+\frac{\dot{r}^{2}}{f(r)}+r^{2}\dot{\phi}^{2}=0,
\end{equation}%
which upon substituting the conserved quantities yields the effective radial
equation 
\begin{equation}
\dot{r}^{2}+V_{\mathrm{eff}}(r)=0,
\end{equation}%
with the effective potential given by 
\begin{equation}
V_{\mathrm{eff}}(r)=-E^{2}+\frac{L^{2}}{r^{2}}f(r).
\end{equation}%
We recall that, circular null geodesics satisfy the conditions 
\begin{equation}
V_{\mathrm{eff}}(r_{\mathrm{ph}})=0,\qquad \left. \frac{dV_{\mathrm{eff}}}{dr%
}\right\vert _{r=r_{\mathrm{ph}}}=0,
\end{equation}%
such that the first condition gives 
\begin{equation}
E^{2}=\frac{L^{2}}{r_{\mathrm{ph}}^{2}}f(r_{\mathrm{ph}}),
\end{equation}%
while the second condition leads to 
\begin{equation}
\left. \frac{d}{dr}\left( \frac{L^{2}}{r^{2}}f(r)\right) \right\vert _{r=r_{%
\mathrm{ph}}}=0.
\end{equation}%
By evaluating the derivative, we obtain 
\begin{equation}
\frac{L^{2}}{r^{2}}\left( f^{\prime }(r)-\frac{2f(r)}{r}\right) =0,
\end{equation}%
such that since $L\neq 0$, this yields the photon sphere condition 
\begin{equation}
2f(r_{\mathrm{ph}})-r_{\mathrm{ph}}f^{\prime }(r_{\mathrm{ph}})=0.
\end{equation}%
Due to the complexity of the metric function is  in the present model, this
equation must generally be solved numerically to determine the photon sphere
radius $r_{\mathrm{ph}}$. The shadow radius is determined by the critical
impact parameter $b=L/E$ that is obtained from $V_{\mathrm{eff}}=0$,  
\begin{equation}
b^{2}=\frac{L^{2}}{E^{2}}=\frac{r_{\mathrm{ph}}^{2}}{f(r_{\mathrm{ph}})}.
\end{equation}%
Therefore, the radius of the shadow observed at infinity is given by 
\begin{equation}
b_{c}=\frac{r_{\mathrm{ph}}}{\sqrt{f(r_{\mathrm{ph}})}}.
\end{equation}%
This corresponds to the boundary between photon trajectories that fall into
the black hole and those that escape to infinity, thereby defining the
apparent shadow. In the Schwarzschild limit ($Q=0$), one recovers the
well-known result $b_{c}=3\sqrt{3}\,M$. For a nonzero electric charge, the
photon sphere radius typically decreases, leading to a smaller shadow size.
The nonlinear parameter $r_{0}$ further modifies the shadow through its
effect on the near-horizon geometry.

In Fig. \ref{F12} we plot the shadow radius for the NED black hole for a
fixed ADM mass but increasing $q$ and compared it with the shadow of the
Reisner-Nordstrom. As the electric charge increases, the photon sphere
radius decreases, leading to a reduction in the shadow size in all cases.
Likewise, increasing the mass $m$ increases the shadow size, consistent with
the chargeless Schwarzschild black hole.

\section{Conclusion}

In this work, we have presented a comprehensive analysis of black hole
solutions arising from Einstein gravity coupled to a nonlinear
electrodynamics (NED) model constructed from a regular electric potential.
The solutions are characterized by three independent parameters, namely the
ADM mass $M$, the electric charge $Q$, and the nonlinear scale $r_{0}$, and
exhibit a remarkably rich structure both geometrically and physically. We
have shown that the spacetime admits four distinct configurations depending
on the values of the dimensionless mass $m$ and charge $q$. These include
black holes with single, double (extremal), and triple (ultra-extremal)
horizons, as well as naked singular solutions. The nature of the central
singularity is controlled by the combination $m-\frac{2}{3}q^{2}$, leading
to either spacelike or timelike singularities. The corresponding Penrose
diagrams were constructed to elucidate the global causal structure in each
case. From a thermodynamic perspective, we derived the Hawking temperature
and heat capacity and identified two critical charge parameters that
determine the stability properties of the system. For $q^{2}<q_{c}^{2}$, the
black hole is thermally unstable, while for $q^{2}>q_{c}^{2}$ a stable phase
emerges within a finite interval of horizon radii bounded by Davies points.
For sufficiently large charge, extremal configurations with vanishing
temperature appear, restricting the stability region to lie between the
extremal radius and the larger transition point. These results demonstrate
that nonlinear electrodynamics introduces nontrivial modifications to the
thermodynamic phase structure beyond the standard electrovacuum case. We
further analyzed null geodesics and showed that the photon sphere is
governed by the condition $2f-rf^{\prime }=0$. The corresponding shadow
radius was computed numerically and compared with the RN solution. We found
that, while both models coincide in the Schwarzschild limit, significant
deviations arise in the strong-field regime. In particular, the NED black
hole may cease to admit a photon sphere beyond a critical charge, in
contrast to the RN case, leading to qualitatively different shadow
properties. These differences become more pronounced at large charge and may
provide a potential observational window through gravitational wave
measurements. Overall, our analysis demonstrates that nonlinear
electrodynamics significantly enriches the geometric, thermodynamic, and
dynamical properties of charged black holes. The combined effects on horizon
structure, stability, and shadow formation, suggest that such models may
offer viable extensions of classical electrovacuum solutions with
potentially observable signatures.

\vspace{1cm} 

\textbf{Conflict of Interest} The authors do not have any
conflict-of-interest.
\vspace{1cm} 

\textbf{Data Availability Statement} This manuscript has no associated data.
Data sharing not applicable to this article as no datasets were generated or
analyzed during the current study.


\begin{thebibliography}{99}
\bibitem{Schwarzschild1916} K. Schwarzschild, On the gravitational field of
a mass point according to Einstein's theory, \emph{Sitzungsber. Preuss.
Akad. Wiss. Berlin (Math. Phys.)}, 189, 196 (1916).

\bibitem{Reissner1916} H. Reissner, \"{U}ber die Eigengravitation des
elektrischen Feldes nach der Einsteinschen Theorie, \emph{Ann. Phys.} 
\textbf{355}, 106 (1916).

\bibitem{Nordstrom1918} G. Nordstr\"{o}m, On the energy of the gravitational
field in Einstein's theory, \emph{Proc. Kon. Ned. Akad. Wet.} \textbf{20},
1238 (1918).

\bibitem{HawkingEllis1973} S. W. Hawking and G. F. R. Ellis, \emph{The Large
Scale Structure of Space-Time} (Cambridge University Press, 1973).

\bibitem{BornInfeld1934} M. Born and L. Infeld, Foundations of the new field
theory, \href{https://doi.org/10.1098/rspa.1934.0059}{\emph{Proc. Roy. Soc. A
} \textbf{144}, 425 (1934)}.

\bibitem{Bardeen1968} J. M. Bardeen, Non-singular general-relativistic
gravitational collapse, Proc. Int. Conf. GR5, 174 (1968).

\bibitem{AyonBeatoGarcia2000} E. Ay\'{o}n-Beato and A. Garc\'{\i}a, The
Bardeen model as a nonlinear magnetic monopole, \href{https://doi.org/10.1016/S0370-2693(00)00357-4
}{\emph{Phys. Lett. B} \textbf{493}, 149 (2000)}.

\bibitem{AyonBeatoGarcia1998} E. Ay\'{o}n-Beato and A. Garc\'{\i}a, Regular
black hole in general relativity coupled to nonlinear electrodynamics, \href{https://doi.org/10.1103/PhysRevLett.80.5056
}{\emph{Phys. Rev. Lett.} \textbf{80}, 505 (1998)}.

\bibitem{Bronnikov2001} K. A. Bronnikov, Regular magnetic black holes and
monopoles from nonlinear electrodynamics, \href{https://doi.org/10.1103/PhysRevD.63.044005
}{\emph{Phys. Rev. D} \textbf{63}, 044005 (2001)}.

\bibitem{Dymnikova2004} I. Dymnikova, Regular electrically charged
structures in nonlinear electrodynamics coupled to general relativity, \href{https://doi.org/10.1088/0264-9381/21/17/009
}{\emph{Class. Quantum Grav.} \textbf{21}, 4417 (2004)}.

\bibitem{FanWang2016} Z. Y. Fan and X. Wang, Construction of regular black
holes in general relativity, \href{https://doi.org/10.1103/PhysRevD.94.124027
}{\emph{Phys. Rev. D} \textbf{94}, 124027 (2016)}.

\bibitem{BalartVagenas2014} L. Balart and E. C. Vagenas, Regular black holes
with a nonlinear electrodynamics source, \href{https://doi.org/10.1103/PhysRevD.90.124045
}{\emph{Phys. Rev. D} \textbf{90}, 124045 (2014)}.

\bibitem{Kruglov2017} S. I. Kruglov, Nonlinear electrodynamics and black
holes, \href{https://doi.org/10.1016/j.aop.2016.12.006}{\emph{Ann. Phys.} 
\textbf{378}, 59 (2017)}.

\bibitem{Hendi2017} S. H. Hendi, Black hole solutions in nonlinear
electrodynamics, \href{https://doi.org/10.1140/epjc/s10052-017-4822-4}{\emph{
Eur. Phys. J. C} \textbf{77}, 296 (2017)}.

\bibitem{Hayward2006} S. A. Hayward, Formation and evaporation of regular
black holes, \href{https://doi.org/10.1103/PhysRevLett.96.031103}{\emph{
Phys. Rev. Lett.} \textbf{96}, 031103 (2006)}.

\bibitem{BambiModesto2013} C. Bambi and L. Modesto, Rotating regular black
holes, \href{https://doi.org/10.1016/j.physletb.2013.03.027}{\emph{Phys.
Lett. B} \textbf{721}, 329 (2013)}.

\bibitem{Mazharimousavi2026} S. H. Mazharimousavi, Regular electric
potential and nonlinear electrodynamics, \emph{Fortschritte der Physik}
(2026), submitted.

\bibitem{Bekenstein1973} J. D. Bekenstein, Black holes and entropy, \href{https://doi.org/10.1103/PhysRevD.7.2333
}{\emph{Phys. Rev. D} \textbf{7}, 2333 (1973)}.

\bibitem{Hawking1975} S. W. Hawking, Particle creation by black holes, \href{https://doi.org/10.1007/BF02345020
}{\emph{Commun. Math. Phys.} \textbf{43}, 199 (1975)}.

\bibitem{Davies1977} P. C. W. Davies, Thermodynamics of black holes, \href{https://doi.org/10.1088/0034-4885/41/8/003
}{\emph{Rep. Prog. Phys.} \textbf{41}, 1313 (1977)}.

\bibitem{Chamblin1999} A. Chamblin et al., Charged AdS black holes and phase
transitions, \href{https://doi.org/10.1103/PhysRevD.60.064018}{\emph{Phys.
Rev. D} \textbf{60}, 064018 (1999)}.

\bibitem{Perlick2015} V. Perlick et al., Influence of a plasma on the shadow
of a spherically symmetric black hole, \href{https://doi.org/10.1103/PhysRevD.92.104031
}{\emph{Phys. Rev. D} \textbf{92}, 104031 (2015)}.

\bibitem{EHT2019a} Event Horizon Telescope Collaboration, First M87 Event
Horizon Telescope Results. I., \href{https://doi.org/10.3847/2041-8213/ab0ec7
}{\emph{Astrophys. J. Lett.} \textbf{875}, L1 (2019)}.

\bibitem{EHT2019b} Event Horizon Telescope Collaboration, First M87 Event
Horizon Telescope Results. VI., \href{https://doi.org/10.3847/2041-8213/ab1141
}{\emph{Astrophys. J. Lett.} \textbf{875}, L6 (2019)}.

\bibitem{ReggeWheeler1957} T. Regge and J. A. Wheeler, Stability of a
Schwarzschild singularity, \href{https://doi.org/10.1103/PhysRev.108.1063}{
\emph{Phys. Rev.} \textbf{108}, 1063 (1957)}.

\bibitem{Zerilli1970} F. J. Zerilli, Gravitational field of a particle
falling in a Schwarzschild geometry, \href{https://doi.org/10.1103/PhysRevD.2.2141
}{\emph{Phys. Rev. D} \textbf{2}, 2141 (1970)}.

\bibitem{KokkotasSchmidt1999} K. D. Kokkotas and B. G. Schmidt, Quasinormal
modes of stars and black holes, \href{https://doi.org/10.12942/lrr-1999-2}{
\emph{Living Rev. Relativity} \textbf{2}, 2 (1999)}.

\bibitem{BertiCardosoStarinets2009} E. Berti, V. Cardoso, and A. O.
Starinets, Quasinormal modes of black holes and black branes, \href{https://doi.org/10.1088/0264-9381/26/16/163001
}{\emph{Class. Quantum Grav.} \textbf{26}, 163001 (2009)}.

\bibitem{Cardoso2009} V. Cardoso et al., Geodesic stability, Lyapunov
exponents and quasinormal modes, \href{https://doi.org/10.1103/PhysRevD.79.064016
}{\emph{Phys. Rev. D} \textbf{79}, 064016 (2009)}.

\bibitem{Wald1993} R. M. Wald, Black Hole Entropy is the Noether Charge, 
\href{https://doi.org/10.1103/PhysRevD.48.R3427}{\emph{Phys. Rev. D} \textbf{
48}, R3427 (1993)}.

\bibitem{IyerWald1994} V. Iyer, and R. M. Wald, Some Properties of Noether
Charge and a Proposal for Dynamical Black Hole Entropy, \href{https://doi.org/10.1103/PhysRevD.50.846
}{\emph{Phys. Rev. D} \textbf{50}, 846 (1994)}.

\bibitem{Rasheed1997} D. A. Rasheed, Non-linear Electrodynamics: Zeroth and
First Laws of Black Hole Mechanics, \href{https://doi.org/10.48550/arXiv.hep-th/9702087
}{arXiv:hep-th/9702087v2}.

\bibitem{Breton2005} N. Breton, Smarr's Formula for Black Holes with
Nonlinear Electrodynamics, \href{https://doi.org/10.1007/s10714-005-0051-x}{
\emph{Gen. Relativ. Gravit.} \textbf{37}, 643 (2005)}.

\bibitem{Dey2004} T. K. Dey, Born-Infeld Black Holes in the Presence of a
Cosmological Constant, \href{https://doi.org/10.1016/j.physletb.2004.06.047}{
\emph{Phys. Lett. B} \textbf{595}, 484 (2004)}.

\bibitem{Hendi2012} S. H. Hendi, Asymptotic Reissner-Nordstr\"{o}m black
holes, \href{https://doi.org/10.1016/j.aop.2013.03.008}{\emph{Ann. Phys.} 
\textbf{333}, 282 (2013)}.
\end{thebibliography}
\end{document}